# MODELING AND EXPERIMENTAL VERIFICATION OF ADAPTIVE 100% STATOR GROUND FAULT PROTECTION SCHEMES FOR SYNCHRONOUS GENERATORS

A Thesis

by

AMIR NEGAHDARI

Submitted to the Office of Graduate and Professional Studies of
Texas A&M University
in partial fulfillment of the requirements for the degree of

MASTER OF SCIENCE

| | |
|---|---|
| Chair of Committee, | Hamid A. Toliyat |
| Committee Members, | Mehrdad Ehsani |
| | Kamran Entesari |
| | Alan Palazzolo |
| Head of Department, | Miroslav M. Begovic |

August 2017

Major Subject: Electrical Engineering

# ABSTRACT

Salient pole synchronous generators as the main component of an electricity generation station should be carefully maintained and their operation has to be monitored such that any damage on them is avoided. Otherwise, the generating station might experience frequent shut downs which results in electricity generation interruptions and high costs associated with repairing and compensation of lack of energy. In this sense, many protective schemes focusing on a variety of synchronous generator faults have already been proposed and are still modified and developed to further enhance the quality of protection.

In this thesis, synchronous generator stator windings to ground fault is studied as one of the most common and crucial faults in these machines. Numerous methods of stator winding to ground fault protection schemes are also reported in the literature. Third harmonic differential voltage and sub-harmonic schemes are studied in this research. A novel adaptive scheme for both methods is modelled and implemented in a comprehensive lab scale set-up where a real generation unit is scaled down including all different components and apparatus. The simulation model is also established based on simultaneous finite element analysis (FEA) and coupled magnetic circuit to assist with system configuration design and parameter selections.

The adaptive scheme is proved to be capable of detecting stator windings to ground faults based on actual experimental data. Finally, the proposed adaptive scheme is compared against other available non-adaptive protection schemes currently used in industrial relays. Several important performance evaluation criteria in protection schemes such as sensitivity and security of operation referred to as reliability are considered. It is shown that the adaptive scheme offers higher reliability than other schemes which emphasizes its credibility and applicability.

# DEDICATION

To my parents

# ACKNOWLEDGEMENTS

I consider myself lucky to get the opportunity of working with my supervisor Prof. Hamid A. Toliyat. I thank him for his continuous support and engagement, his valuable time, his commitment to excellence in research, his patience and his sense of humor.

I am grateful to my committee members Prof. Mehrdad Ehsani, Dr. Kamran Entesari and Prof. Alan Palazzolo. I learnt a lot from attending their courses and advising during the course of study at Texas A&M University.

I would also like to thank ECE department staff – Mrs. Tammy Carda, Melissa Sheldon, Anni Brunker and Katharine Bryan who have always been very pleasant and helpful.

I express my deep gratitude to my colleagues at EMPE lab and EPPE group at ECEN department for their help in conducting experiments, building test setups and technical discussions. I would like to thank my co-worker Dr. Khaled Al Jaafari for his help in this work. I also thank all my other labmates, Drs. Abdulkadir Bostanci, Vivek Sundaram, Yateendra Deshpande, Siavash Pakdelian, Jae-Bum Park, Matthew Johnson, Hussain Hussain and Bahar Anvari, Matthew Gardner, Morteza Moosavi, Ajay Kumar Morya, Farid Naghavi and Niloofar Torabi who help me feeling comfortable in the lab.

I am thankful of being lucky to cooperate with Dr. Nader Safari-Shad and Russel Franklin of Alliant Energy on this project. I really appreciate their patience and continuous advices during more than two years that was spent on this research project.

Last but not the least, I would like to thank my parents for their love, patience, selfless support and encouragement over the years. This work would have not been possible without their support.

# CONTRIBUTORS AND FUNDING SOURCES

**Contributors:**

*Part 1, faculty committee recognition*

This work was supervised by a thesis committee consisting of Professor Hamid A. Toliyat [advisor] and Professors Mehrdad Ehsani and Kamran Entesari of the Department of Electrical and Computer Engineering and Professor Alan Palazzolo of the Department of Mechanical Engineering.

*Part 2, student/collaborator contributions*

All work for the thesis was completed by the student, in collaboration with Dr. Nader Safari-Shad and Mr. Russel Franklin.

**Funding Sources:**

This work was made possible by partial financial support received from Alliant Energy Inc.

# NOMENCLATURE

| | |
|---|---|
| $C_s$ | Stator Windings Coupling Capacitance |
| $C_t$ | Connected Apparatus Shunt Capacitances |
| $E_3$ | Third Harmonic Voltage Source |
| T | Time |
| $V_{P3}$ | Phase Third Harmonic Voltage |
| $V_{N3}$ | Neutral Third Harmonic Voltage |
| $Z_g$ | Grounding Impedance |
| $\rho$ | Phase and Neutral Third Harmonic Ratio |
| $\nu$ | Random White Noise |
| e | Random White Noise |
| $\rho_0$ | Mean of Random White Noise |
| $\Pi_0$ | Variance of random White Noise |
| E | Expected Value |
| $\delta$ | Kronecker Delta |
| $P_{(t)}$ | Estimation Error Covariance |
| $K_{(t)}$ | Kalman Filter Gain |
| $J_{AO}$ | A64G2 Operate Qauntity |
| $J_{AR}$ | Restraint Quantity |
| $\beta$ | Sensitivity Factor |
| $R_n$ | Neutral Grounding Resistor |
| $R_f$ | Fault Resistor |
| $C_p$ | Parasitic Capacitors |
| $L$ | Sampling Window |
| $N$ | Neutral Grounding Transformer Turns Ratio |
| $R_{BPF}$ | Band Pass Filter Resistance |
| $X_c$ | Total Machine Capacitance |
| $R_s$ | Insulation Resistance |
| $V_N$ | Neutral Voltage |
| $I_N$ | Neutral Current |
| $V_s$ | Sub-harmonic Voltage Source |
| $H_{1(s)}$ | Transfer Function from $V_s$ to $I_N$ |
| $H_{2(s)}$ | Transfer Function from $V_s$ to $V_N$ |
| $H_{3(s)}$ | Transfer Function from $I_N$ to $V_N$ |
| $\tau_0$ | Machine Time Constant |
| $z$ | Discrete Time Complex Variable |
| $T$ | Sampling Period |
| $\Theta_0$ | Gaussian Random Parameter Vector |

# TABLE OF CONTENTS

# LIST OF FIGURES

# LIST OF TABLES

# 1 INTRODUCTION

## 1.1 Background

Synchronous generator stator winding to ground faults are a common cause of generator failure. It mainly occurs due to two major reasons. First, the mechanical damages to the ground insulation and second the insulation burning. The ultrahigh currents flowing through the windings could cause severe damages to the windings as shown in Fig. 1.1.

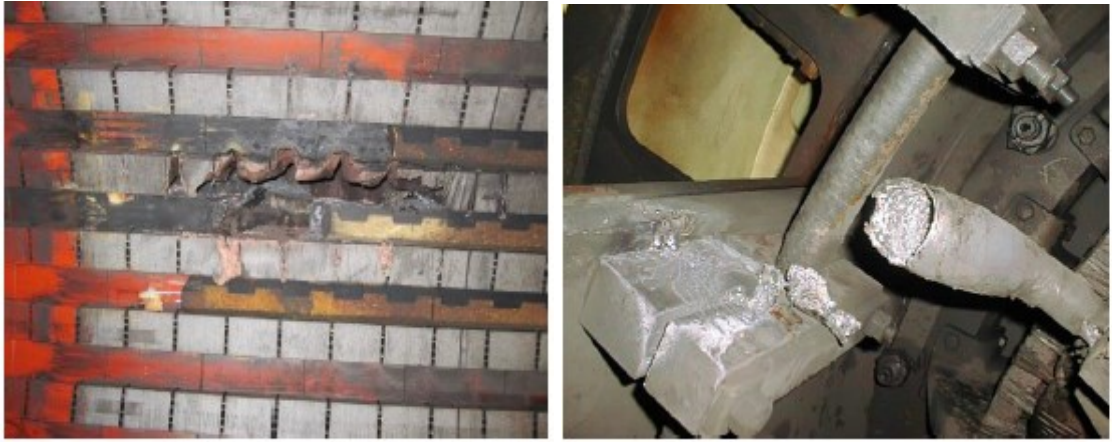

Fig. 1.1: Stator windings insulation failure.

There are many stator ground fault protection schemes reported in the literature [1-14]. Each uses different theoretical basis and possesses strengths and weaknesses. For example, some only detect faults occurring at certain portion of the winding and are unable to cover the entire length of the winding or some might misoperate when there is a disturbance in the actual operation of the system. Among all these different methods, the followings are more popular such that they have been utilized or currently are used in industrial relays. The neutral fundamental frequency overvoltage scheme (59N) generally performs well in detecting faults along 90% - 95% of the stator winding [15, 16]. The protection of remaining 5% - 10% of the winding toward the neutral point is challenging due to low induced neutral voltage. The third harmonic differential voltage scheme (64G2) is based on third harmonic voltage distribution along the stator winding specially on neutral and terminal ends of the winding. The ratio of third harmonic voltage on neutral and terminal is used as a signature which alters while faults happen in the system. The sub-harmonic voltage injection scheme (64S) is believed to protect 100% of the stator winding [17, 18, 19]. However, its application depends heavily on accurate identification of the total capacitance to ground of the generator stator windings, iso-phase bus work and delta-connected windings of the step-up transformer [20]. To protect 100% of the stator windings, the third-harmonic schemes are used in conjunction with the 59N scheme [21]. Two most notable third-harmonic schemes are the undervoltage (27TN) scheme and the third-harmonic differential voltage (59THD) scheme.

**1.2  Different Fault Types**

The grounding method of the synchronous generators and the way that stator windings to ground fault occurs is categorized into various types of faults, [22-29]. Low resistance grounding is referred to those faults in which the winding to ground is a low resistance path and hence it might lead to high fault currents. Single point grounding is where only one point along the stator windings is shorted to the ground. Usually, it is easier to detect and locate these type of faults. On the other hand, multiple point grounding means several point of different stator winding phases and belonging to different synchronous generator could be shorted. Unlike single point grounding, detection and locating the faults would be more challenging in this case. High impedance grounding which is the case considered in this thesis is where the neutral connection of the stator windings is connected to an impedance via a neutral voltage transformer that is grounded. The faults currents are relatively low and hence high precision detection schemes are required. In hybrid approach both high and low resistance grounding paths are hired and they are often switched with each other to attain the safest operation of the machine.

## 1.3 Literature review

Generally as part of the fulfillment of the acceptable protection of synchronous generators against various faults, several protection schemes have been proposed so far each possessing advantages and disadvantages as well as using different signals and methods as a basis in their algorithm. Bellow some of the most highlighted ones are described:

• Phase undervoltage (27): This scheme sends an alarm when the phase terminal fundamental voltage drops below a certain value. It is not capable of detecting faults closer to neutral terminal of the winding as in these cases the phase voltage will be increased.

• Neutral third harmonic undervoltage (27TN): This scheme utilizes the neutral third harmonic voltage as an indicator of fault and sends alarm whenever this signal falls under a certain limit. While the fault is located on the phase terminal, the third harmonic voltage on the neutral goes up and this scheme fails to detect.

• Instantaneous neutral overcurrent (50N): This scheme continuously monitors the current flowing through the neutral connection of the stator windings. As soon as the fault occurs, this current will be increased compared to non-faulty operation and if the threshold settings are chosen appropriately the alarm can be sent out.

• Third harmonic voltage differential ratio (59D): This scheme uses both the neutral and phase terminal third harmonic voltages as the input and the ratio of them is a fault indicator. Due to synchronous generators third harmonic characteristic described later in the next section, the value of this ratio can be estimated under different fault conditions. The first chapter of this thesis presents an improved version of third harmonic differential voltage ratio.

• Low frequency injection (64S): This scheme is claimed to cover the entire length of the stator winding or in other words provides 100% stator windings to ground fault protection. A low frequency signal (in order not to interfere with the fundamental frequency) is injected to the neutral circuit. Whenever the fault happens, the equivalent impedance of the machine referred to the secondary side of the neutral transformer is reduced and a higher subharmonic current flows into the neutral circuit which is used as a fault indicator. The second chapter of this thesis discusses an adaptive subharmonic injection scheme which relieves the disadvantages of the conventional subharmonic schemes.

## 1.4  Objectives and overview

The third harmonic differential voltage and subharmonic injection are studied respectively in this thesis. In some synchronous machines, the third harmonic voltages at the generator neutral and terminal vary due to generator loading, power factor variation and system disturbances. These variations make the third-harmonic schemes insecure resulting in false alarms or misoperations leading to costly loss of production and testing of the generator. The insecurity issues associated with the conventional third-harmonic schemes are examined in [30] using simulations of field data from five large industrial generators. To improve security, a new adaptive third-harmonic differential voltage scheme based on Kalman adaptive filtering (KAF) is proposed in [30]. The improved security is then shown using disturbance simulations of field data and actual generator misoperation events. The present thesis reports on modelling and experimental verification of the scheme proposed in [30]. To this end, a laboratory scale synchronous generator setup with the help of a detailed FEA method of the overall system is built to closely emulate the third-harmonic voltage fault characteristics of industrial synchronous generators. The data from fault studies has so far established the dependability of the adaptive scheme, i.e., correct operation due to faults.

## 2 THIRD HARMONIC DIFFERENTIAL VOLTAGE SCHEME

The third harmonic differential voltage scheme can be mentioned as one of the most popular stator windings ground fault protection methods for synchronous generators; However, this scheme is not capable of detecting faults happening all over the winding length, by combining it with other schemes (usually neutral fundamental undervoltage scheme), 100% stator winding coverage can be achieved.

The synchronous generator third harmonic distribution model is the core sign in third harmonic differential voltage scheme developments. Third harmonic voltage is uniformly distributed along the stator winding during healthy (no fault) operation of the machine and will be redistributed based on a known characteristic after the fault occurs. Fig. 2.1 explains the split capacitance model of the generator stator windings. In this figure, "Cs" represents the stator winding capacitance and "Ct" represents all the shunt capacitances at the generator terminals including the capacitance of the primary windings of the connected transformers. Using this approximate model, half of the parallel distributed capacitances along the winding are transferred to the neutral and the other half is assumed to belong to the terminal side as illustrated in Fig. 2.1. In Fig. 2.1, total third harmonic voltage of the winding (E3) is divided between the neutral and the terminal by the coefficient 'α' where $0 \leq \alpha \leq 1$.

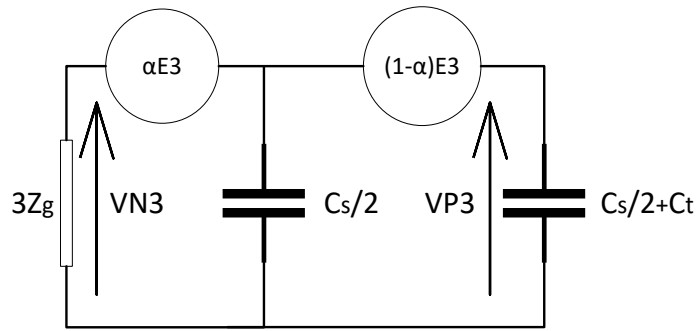

Fig. 2.1: Split capacitance model of SG, adapted from [39].

Splitting the capacitances causes the harmonic voltages to be divided between the terminals and the neutral as illustrated in Fig. 2.2. When the ground faults happen close to the neutral, the capacitive impedances close to the neutral and the neutral impedance "3Zg" are shorted and the third harmonic voltage in the same area will be reduced.

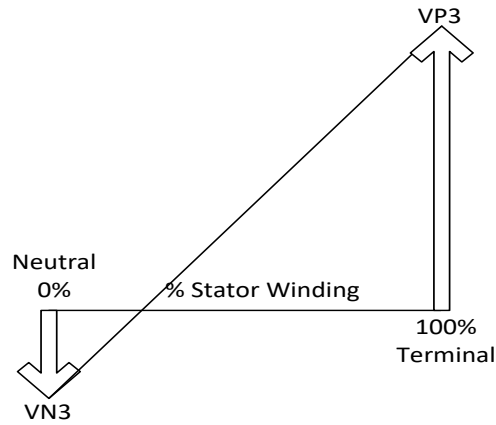

Fig 2.2: Voltage distribution from neutral to terminal of SG, adapted from [39].

Therefore, all harmonic voltages will appear at the terminals i.e.,

$$V_{N3} \approx 0 \qquad V_{P3} \approx E3 \tag{2.1}$$

In contrast, when the ground faults happen close to the terminal, the capacitive impedances close to the terminal is shorted and the 3rd harmonic voltage in the same area will be reduced. Therefore, all harmonic voltages will appear at the neutral i.e.,

$$V_{P3} \approx 0 \qquad V_{N3} \approx E3 \tag{2.2}$$

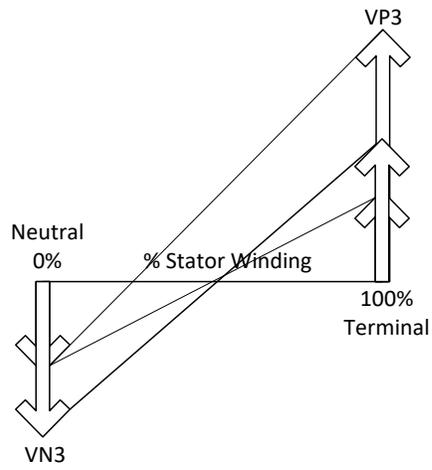

(a)

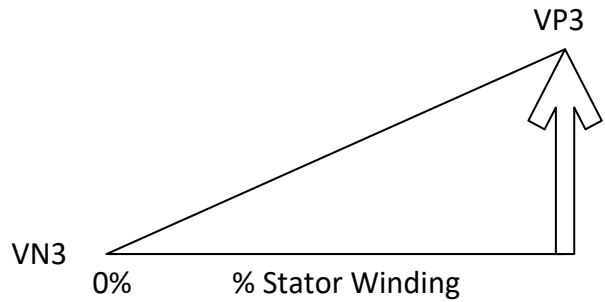

(b)

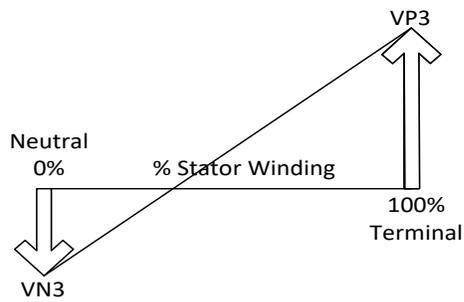

(c)

Fig 2.3: SG third harmonic model under: (a) normal operation, (b) fault at the neutral end, and (c) fault at 30% from the neutral end, adapted from [39].

The third harmonic voltage distribution during normal and fault conditions is shown in Fig. 2.3. Note that neutral faults are of higher importance since the terminal faults can be reliably detected by the 59N fundamental voltage scheme.

The changes in VN3 and VP3 during normal and ground fault conditions is captured by introducing the so-called third harmonic voltage ratio ($\rho$ = VN3/VP3). It is the changes in this parameter that is at the core of the design of many commerical relay operating algorithms, [30, 31].

In the next section, the proposed adaptive 64G2 scheme is explained.

## 2.1 Adaptive 64G2 Scheme theory

*a) System state space model and the proposed Kalman Adaptive Filtering (KAF)*

The key feature in this scheme is the variable third harmonic ratio which adapts itself during different operating conditions of the synchronous machine. Variable "t" stands for discrete-time index in all equations. Modeling the ratio as a variable parameter is necessary for applicability of Kalman filtering theory, [32]:

$$\rho(t+1) = \rho(t) + v(t), \qquad \rho(0) = \rho_0 \qquad (2.3)$$

$$V_{N3}(t) = V_{P3}(t)\rho(t) + e(t) \qquad (2.4)$$

Here, $\rho(t)$ is the third harmonic voltage ratio, $\rho_0$ is an initial ratio having an assumed Gaussian distribution with mean and variance equal to $\hat{\rho}_0$ and $\Pi_0$, $v(t)$ and $e(t)$ are random white noises and $V_{N3}$ and $V_{P3}$ represent third harmonic voltage magnitudes at neutral and terminal of the stator winding, respectively. Furthermore:

$$E\{v(t)v(s)\} = Q(t)\delta(t,s), \quad E\{e(t)e(s)\} = R(t)\delta(t,s) \qquad (2.5)$$

where $\delta(t,s) = 1$ if t=s, otherwise $\delta(t,s) = 0$. The reason noises were added to the above model is to account for any probable disturbance and model inaccuracies which might create error.

The KAF model describing the minimum-variance estimate of the ratio is given by:

$$\hat{\rho}(t) = \hat{\rho}(t-1) + K(t)\Delta(t) \quad (2.6)$$

where,

$$\Delta(t) = V_{N3}(t) - V_{P3}(t)\hat{\rho}(t-1) \quad (2.7)$$

and,

$$K(t) = P(t)V_{P3}(t)R^{-1}(t) \quad (2.8)$$

$$P(t) = \frac{P(t-1)R(t)}{R(t) + P(t-1)V_{P3}^2(t)} + Q(t) \quad (2.9)$$

where, P(t) is the estimation error covariance, and K(t) is the Kalman filter gain.

*b) Adaptive 64G2 scheme fault detection criterion*

To arrive at a reliable ground fault detection, [9] introduces two newly defined operate and restraint quantities denoted by $J_{AO}(t)$ and $J_{AR}(t)$:

$$J_{AO}(t) = \begin{cases} 0, & \text{if } 1 \leq t \leq L \\ \sum_{i=t-L}^{t} |v(i)|^2, & \text{if } t > L \end{cases} \quad (2.10)$$

$$J_{AR}(t) = \begin{cases} \sum_{i=1}^{t} V_{N3}^2(i), & \text{if } 1 \leq t \leq L \\ \sum_{i=t-L}^{t} V_{N3}^2(i), & \text{if } t > L \end{cases} \quad (2.11)$$

where "L" is a positive integer representing the window width over which the third harmonic voltage ratio is learned and $J_{AO}$ is inhibited. The trip equation for the new adaptive 64G2 scheme is defined by:

$$J_{AO}(t) > \beta J_{AR}(t) \tag{2.12}$$

where "β" is a sensitivity factor. It is important to note that the new scheme called A64G2 will detect a stator ground fault whenever (2.12) is true over a window of L sample points.

To summarize, this scheme utilizes the provided third harmonic voltages on both neutral and terminal of the stator windings in conjunction with the power factor as input. It feeds them through KAF where the residual signal Δ(t) is generated then the adaptive fault detector decides whether to send an alarm, trip or take no action. The operational block diagram of the system is shown in Fig. 2.4.

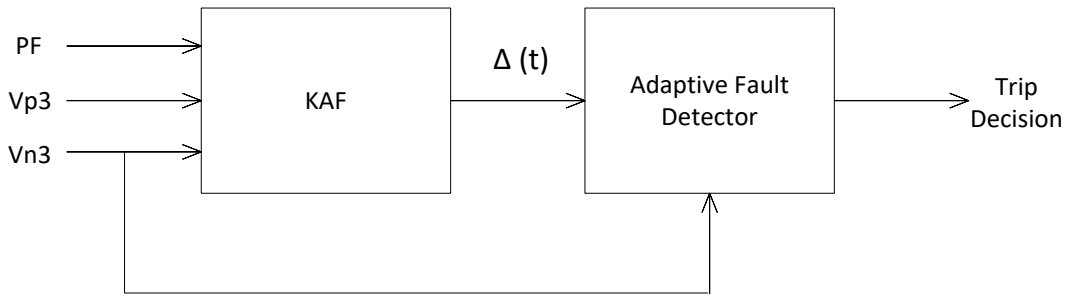

Fig 2.4: Block diagram of the A64G2 operation scheme, adapted from [39].

## 2.2 System Modelling and Simulation

In order to study SG winding faults and their effects on the generator third harmonic voltages a two dimensional (2D) transient FEA model of the salient pole synchronous generator shown in Fig. 2.5 is considered.

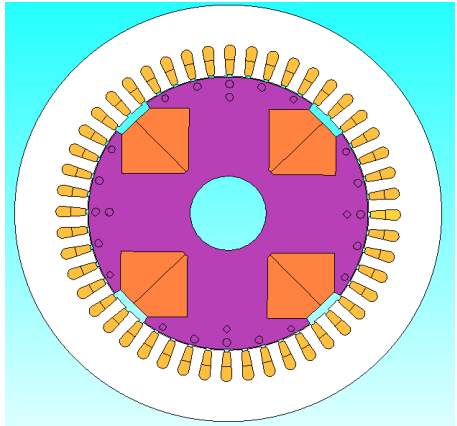

Fig 2.5: FEA model of the SG's stator and rotor showing double layer short pitched windings, adapted from [39].

The SG specifications that later is used in experimental study is given in Table 2.1.

Table 2.1: SG parameters.

| Rated power | 5 kVA |
|---|---|
| Number of Poles | 4 |
| Rated voltage | 240 V |
| Rated current | 12 A |
| Power factor | 0.8 |
| Number of winding layers per stator slot | 2 |
| Number of turns per coil | 6 |
| Number of coils per phase | 4 |

| Number of stator slots | 48 |
|---|---|
| Air gap length | 0.5 mm |
| Rotor outer radius | 87 mm |
| Rotor inner radius | 23.75 mm |
| Stator outer radius | 133 mm |
| Fault resistor ($R_f$) | 50 Ω |
| Neutral resistor (R'n) | 350Ω |
| Total SG parasitic capacitors to ground ($15*C_p$) | 7.5 μF |

The circuit model of the machine including stator windings, distributed coupling capacitors and fault resistor is shown in Fig. 2.6.

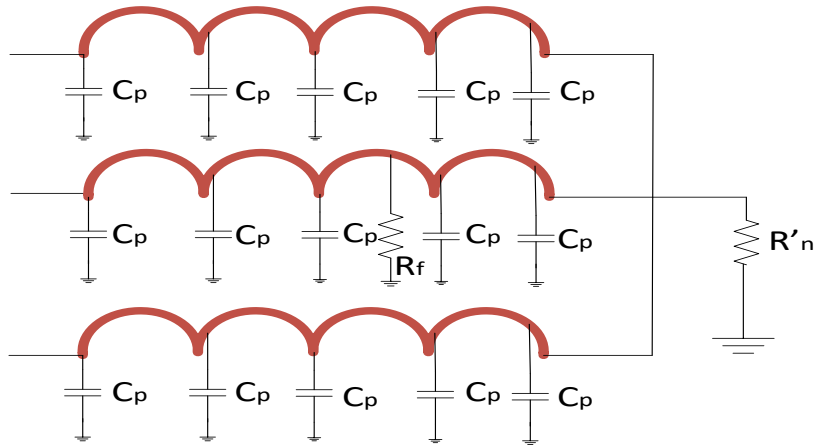

Fig 2.6: Circuit model coupled with FEA simulation, adapted from [39].

Due to very low stator windings to ground capacitors in the lab scale SG compared to industrial SG's, fault characteristics cannot be emulated. In order to resolve this issue, some external capacitors are required to be added uniformly along the stator windings.

The quality of distribution and magnitudes of the external capacitors mainly depend on machine specifications and ratings. In this system, five uniformly distributed 0.5 µF capacitors are added to each phase as shown in Fig. 2.6 to make sure that the fault characteristics is realized. The three phase and neutral output voltages of the machine operating at rated conditions were collected and FFT was performed on them to extract third harmonic levels. The expected third harmonic characteristics during pre-fault and post-fault conditions are shown in Figs. 2.7 and 2.8. Note that hard neutral and hard terminal faults refer to faults caused at 0% and 100% of the windings, respectively.

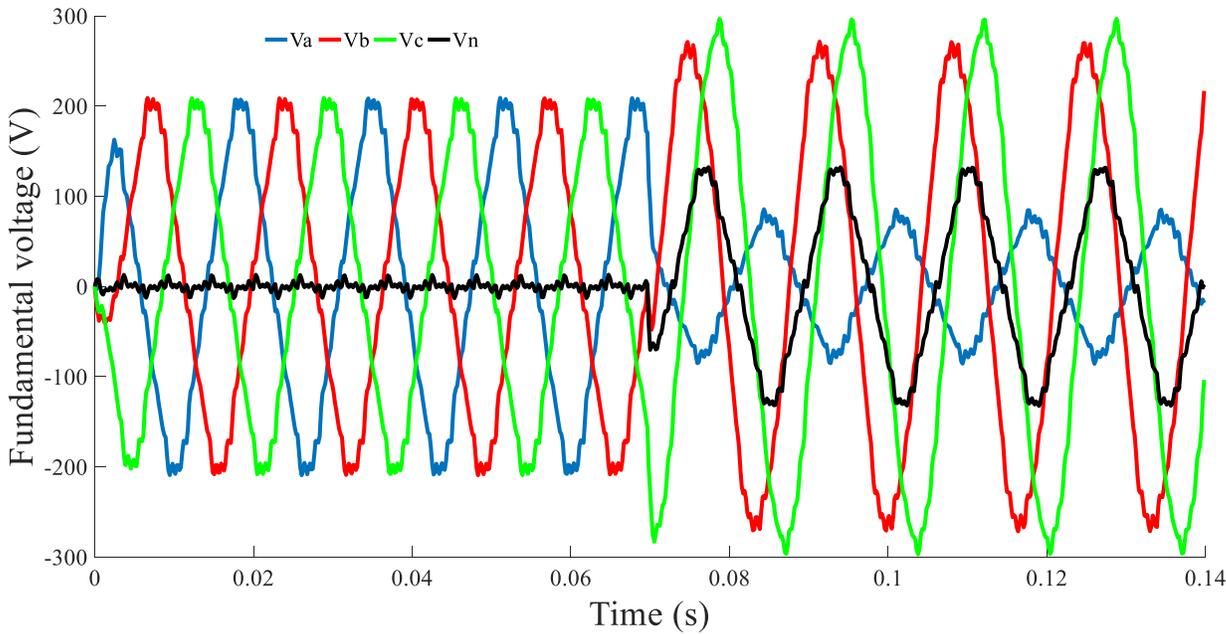

(a)

Fig. 2.7: Three phase and neutral voltages before and after hard terminal fault (at phase A): (a) Fundamental voltage, and (b) Third harmonic voltage based on FEA simulation, adapted from [39].

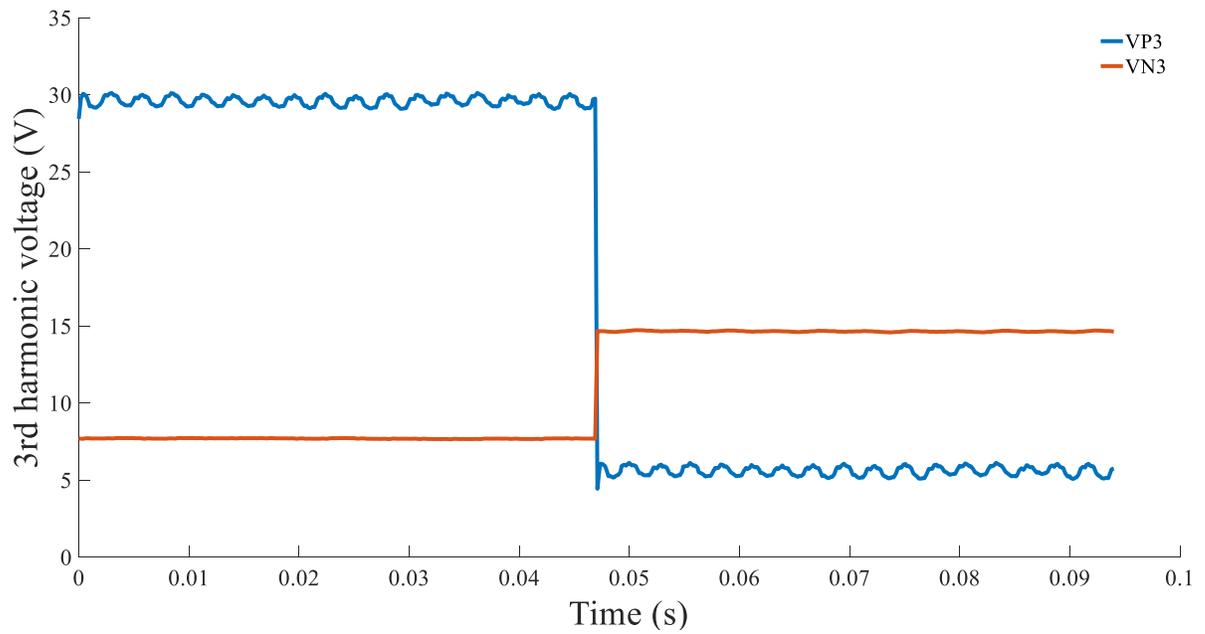

(b)

Fig. 2.7 Continued.

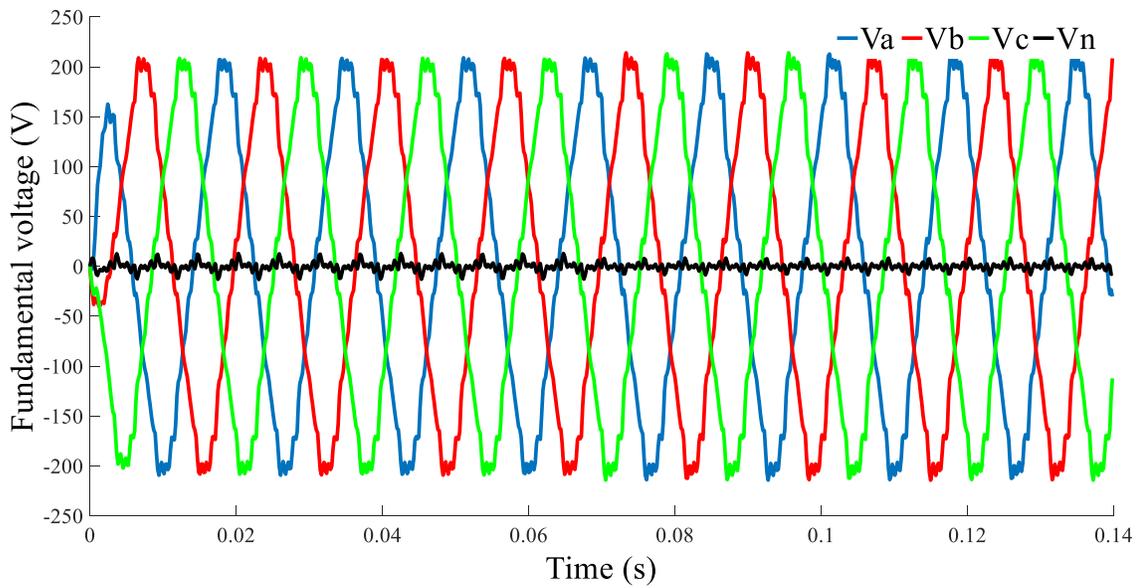

(a)

Fig. 2.8: Three phase and neutral voltages before and after hard neutral fault: (a) Fundamental voltage, and (b) Third harmonic voltage based on FEA simulation, adapted from [39].

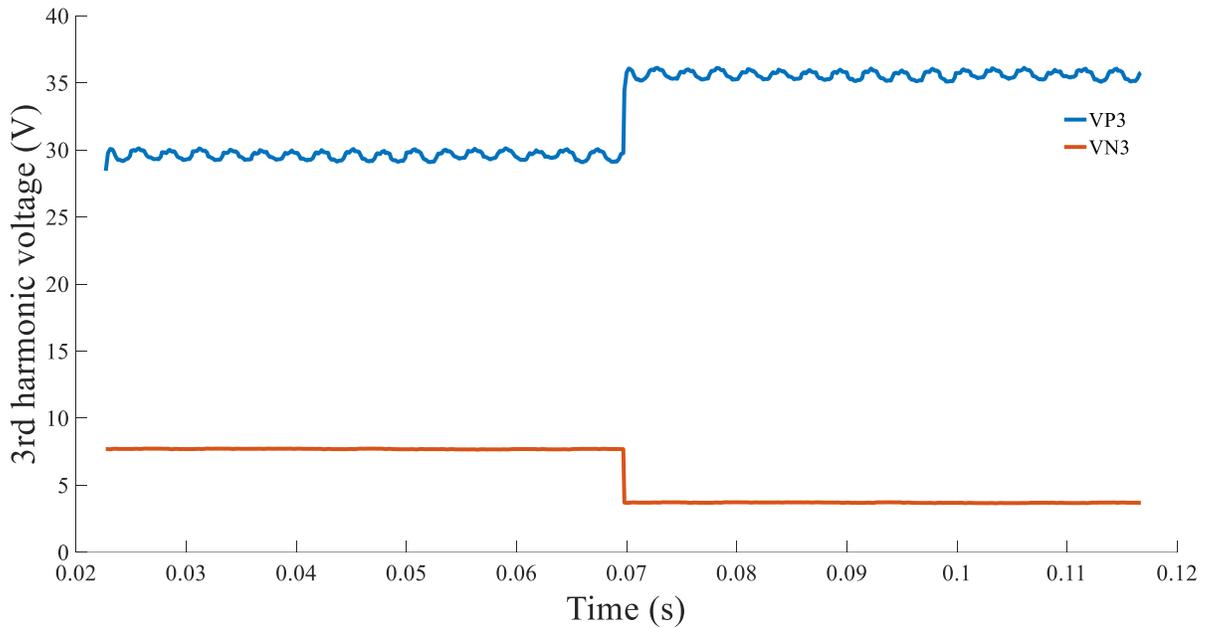

(b)

Fig. 2.8 Continued.

In the 2D FEA that was performed, the skew effect was neglected which in reality it decreases the 3$^{rd}$ harmonic level. The lower 3$^{rd}$ harmonic level observed during experiment than FEA simulation arises from this fact.

It should be mentioned that whenever three phases' third harmonic levels is mentioned, the summation of third harmonics of all three phases is meant. Since they are in phase with respect to each other, this summation is converted to a scalar summation. In Figs. (2.7-b), and (2.8-b) $V_{P3}$ and $V_{N3}$ are primary voltages of phase potential transformers (PTs) and neutral grounding transformer, respectively.

To induce a fault, the stator winding in one of the phases is shorted to ground through the fault resistor (50Ω) at t=0.069s (sample point 270). All simulation data were fed into the adaptive scheme to evaluate its performance and the results are shown in Figs. 2.9 and 2.10.

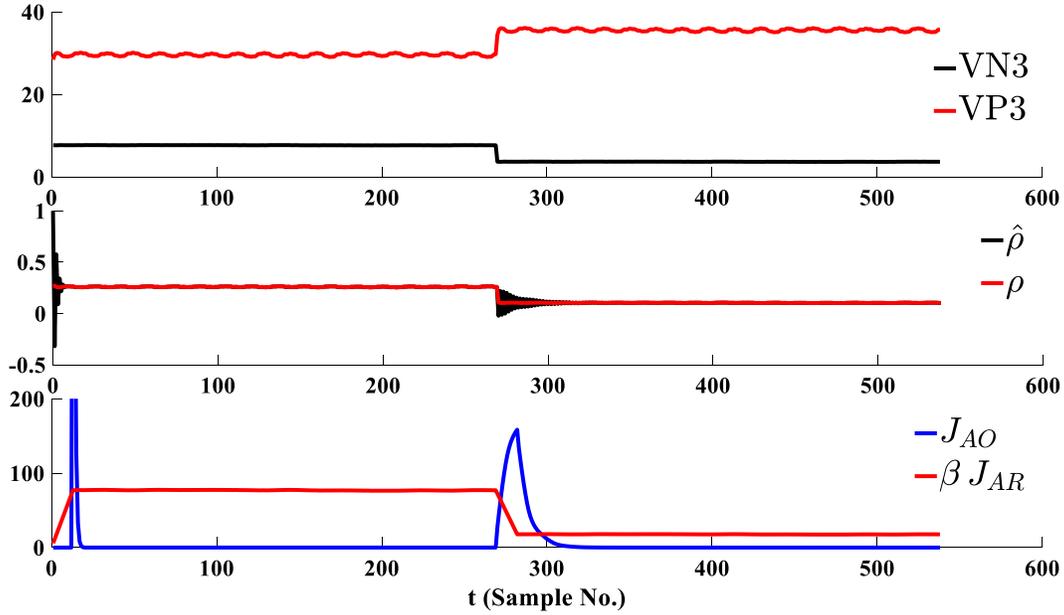

Fig 2.9: Simulation responses of adaptive 64G2 scheme to hard neutral fault, adapted from [39].

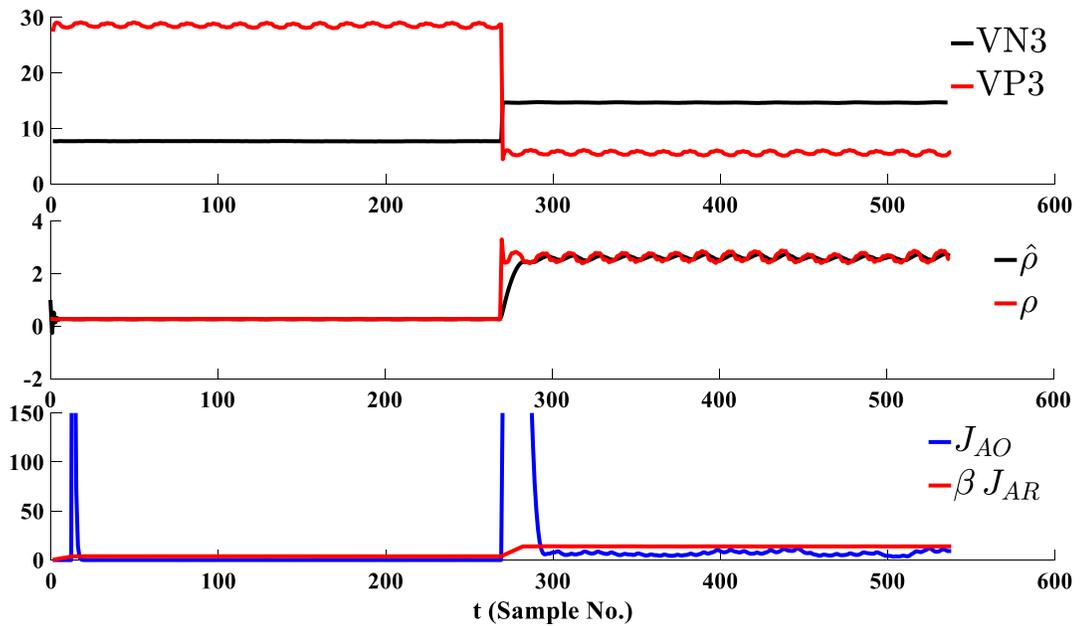

Fig 2.10: Simulation responses of adaptive 64G2 scheme to hard terminal fault, adapted from [39].

The value of parameter "β" in adaptive scheme is chosen to be 0.005 to assure not making the adaptive scheme too sensitive whereas "$\beta$" in large generators used in electric power plants was chosen to be 0.25, [30].

Accordong to Figs. 2.9 and 2.10, whenever the operating signals cross over the restraint signals and stay above for at least L samples (L=12), the fault is detected. The sharp cross over during the learning period is too short (only two data point) and will not cause a trip signal. The so called blind zone is that fraction of the winding along which expected fault characteristic is not emulated. The blind zone of this SG based on simulation results shown in Fig. 2.11 is estimated to be from 44% to 56% of the winding length.

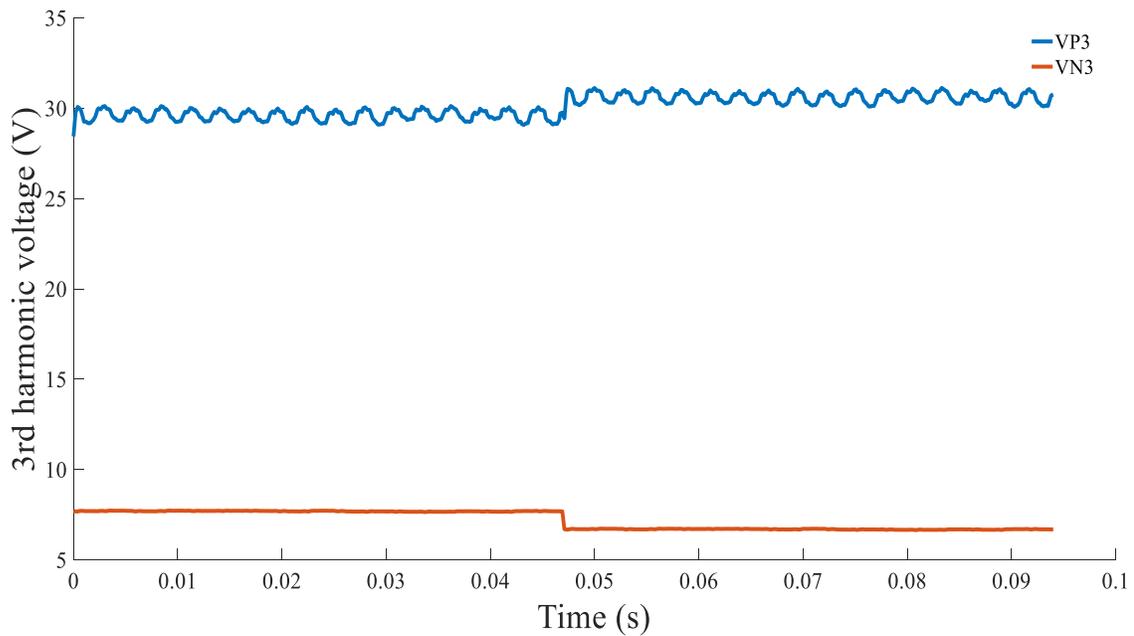

(a)

Fig 2.11: SG fault characteristics at tabs on (a) 44%, (b) 50%, (c) 56% of the winding, adapted from [39].

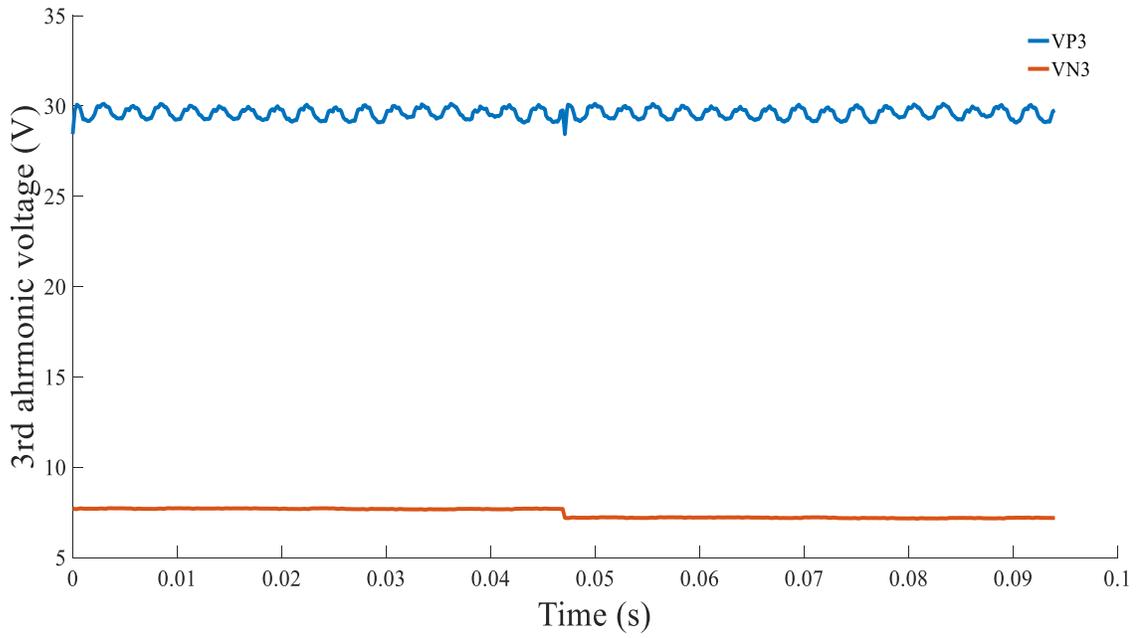

(b)

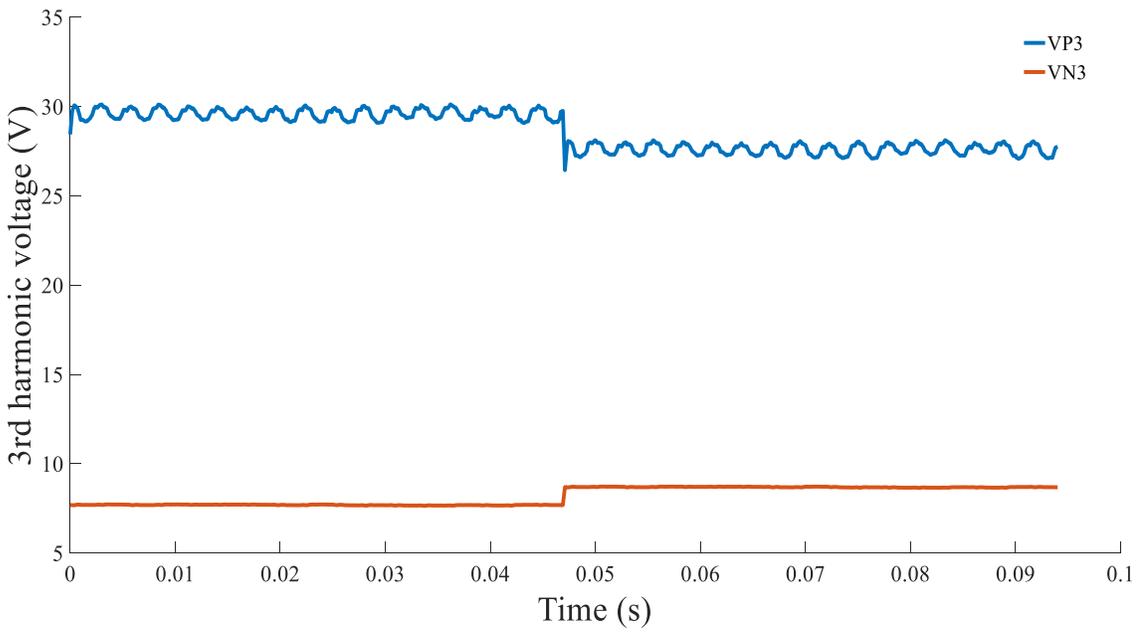

(c)

Fig 2.11 Continued.

## 2.3 Experimental Set-up and Results

One of the very distinctive features of this work is the construction of a lab prototype emulating an electric power plant generator. In practice, there are high costs associated with collection of real time data on large scale industrial generators. Moreover, unwillingness of power plant managers to conduct tests is a major obstacle to obtaining a rich set of field data. To overcome this issue, an experimental test bed containing the required synchronous machine was modified to model a real electric generation unit which consists of transmission and distribution transformers, stator windings terminal potential transformers (PTs), data collection unit, load, etc. An overview of such a system is shown in Fig. 2.12.

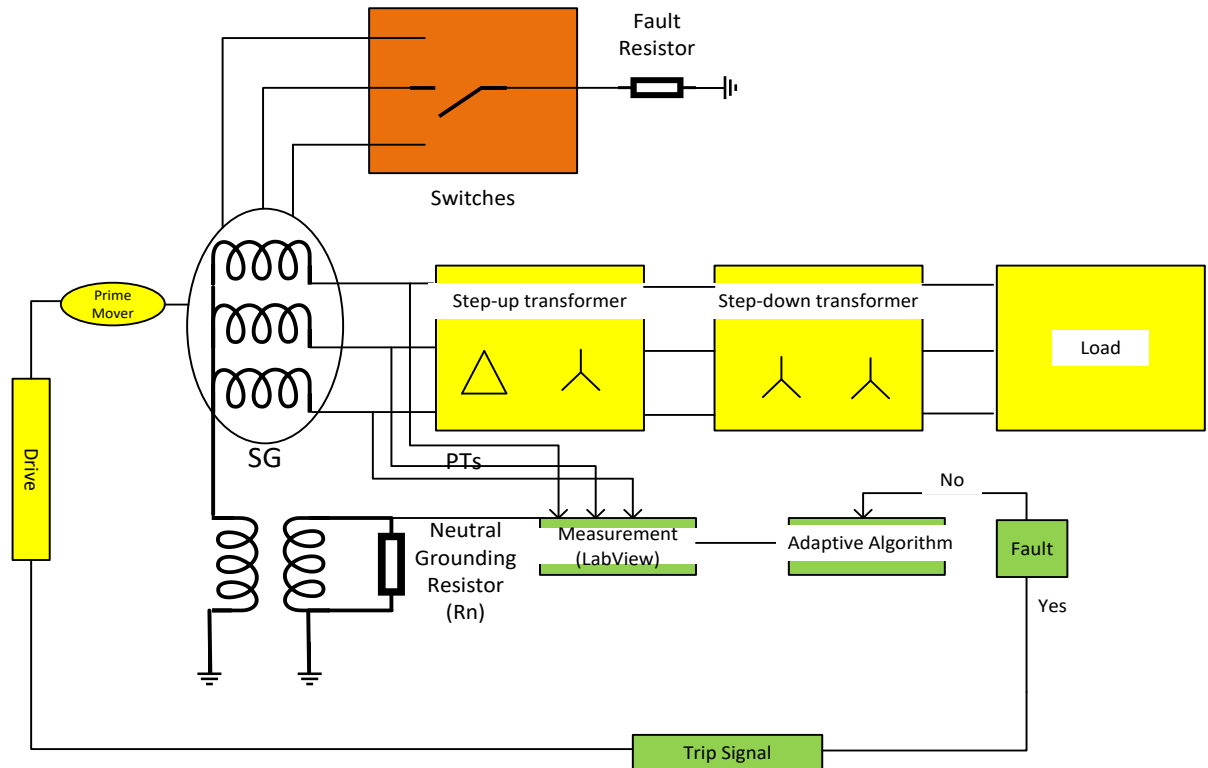

Fig 2.12: Overall system schematic, adapted from [39].

The overall constructed system is intended to model a real generating station unit including all different subsystems such as power system, control and protection, and fault/disturbance injector subsystems. Each part is described below:

• Power system: Synchronous generators, step-up transformers, transmission lines, step-down transformers, distribution lines and loads are the major component of the power system. For the lab setup these subsystems are shown in Fig. 2.12. A lab scale SG was chosen for this study representing the large scale SGs in the power system. Several parameters of this SG can be found in Table 2.1. The rotor and stator were disassembled to rewind the stator windings in order to add the external taps on it. Fig. 2.13 shows the linear diagram of the winding arrangement for three phase windings.

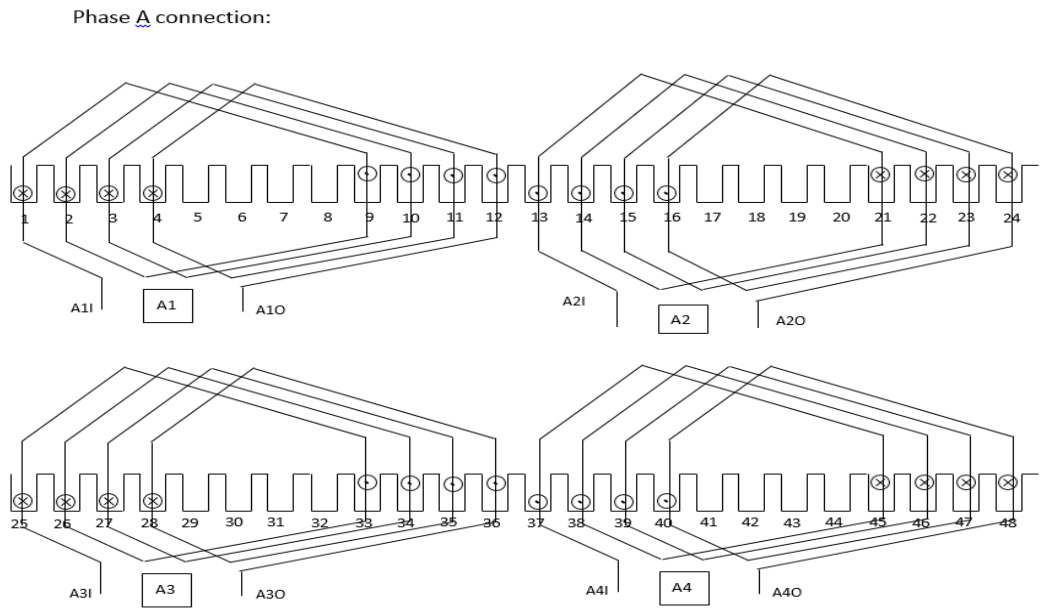

(a)

Fig 2.13: Stator winding configuration a) Phase A, b) Phase B, and c) Phase C.

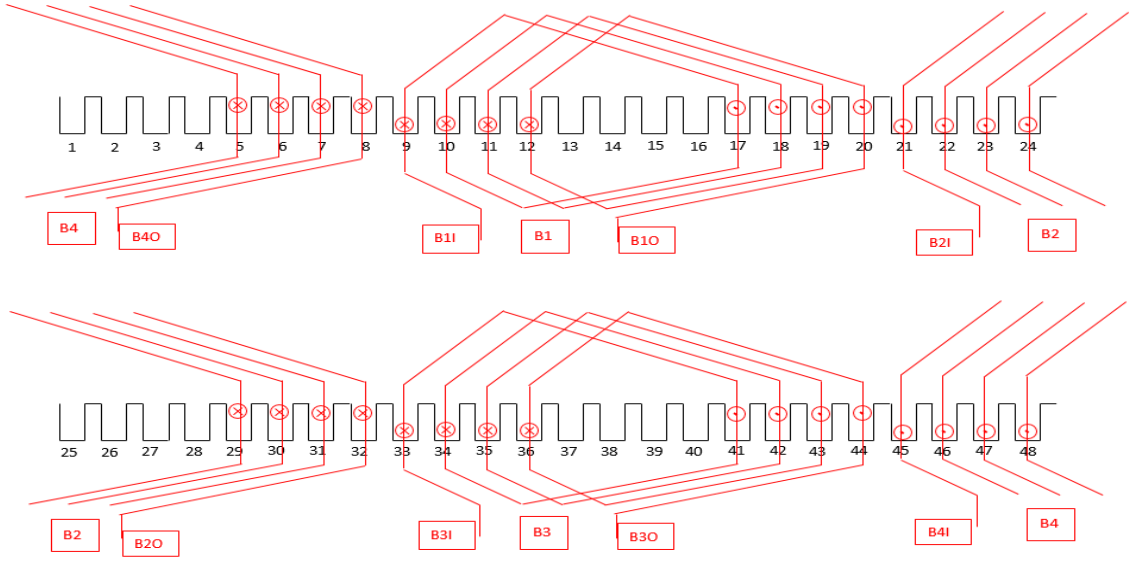

(b)

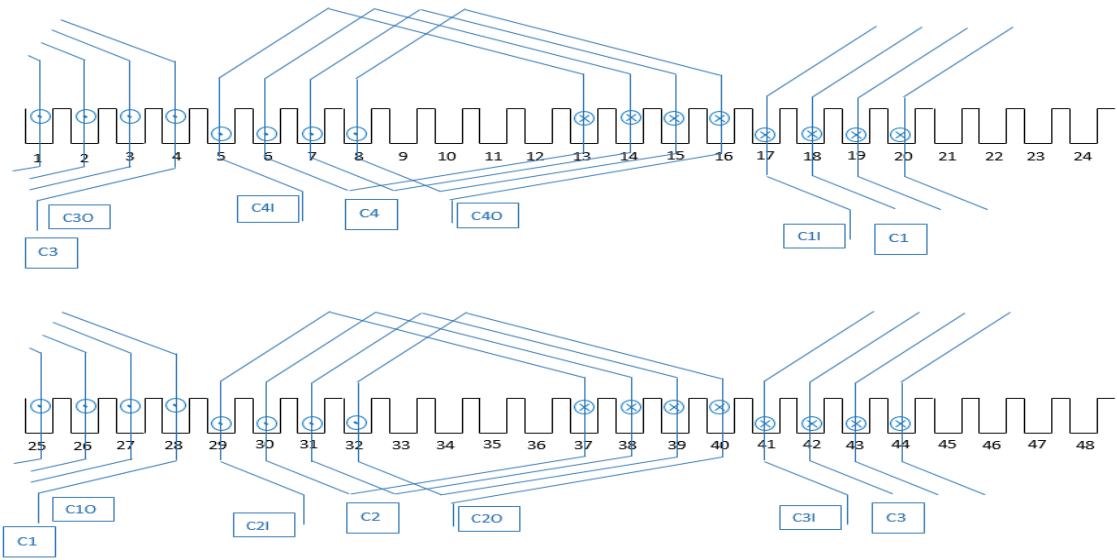

(c)

Fig 2.13 Continued.

The final wirings coming out of the stator winding terminals are shown in Fig. 2.14. The three phase star point neutral connection, phase terminals, field excitation, end coil connections and the external taps are the accessible points.

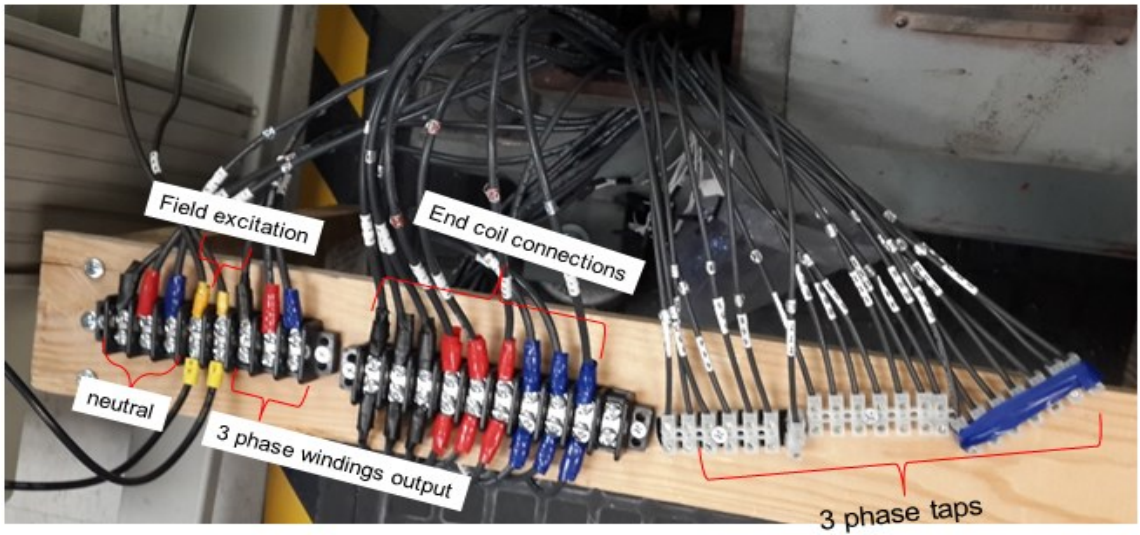

Fig 2.14: Accessible stator winding connections.

The neutral grounding transformer shown below is intended to isolate the stator winding neutral point and the measurement interface. It is also responsible for reflecting the neutral resistor with higher coefficient to the primary side to assure high impedance grounding feature. It possesses two 220V windings on the primary side as well as two 120V windings on the secondary side.

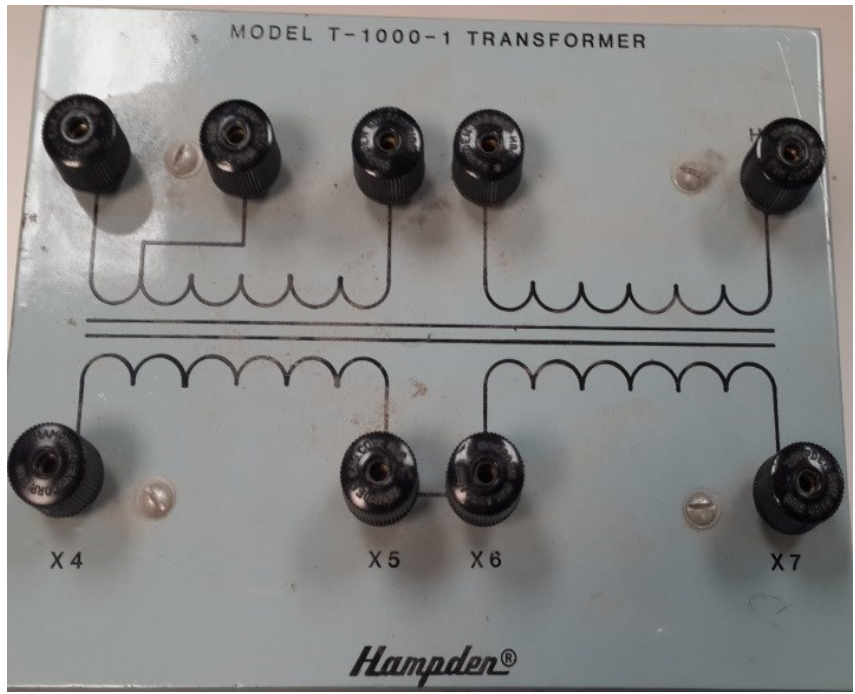

Fig 2.15: Neutral grounding power transformer.

The step up transformer is a delta-wye connected transformer whose connections and ratings are shown below. The primary delta connection is to avoid harmonic currents of the machine to be flowed toward the load and grid. The step down transformer connections and specifications are also shown below.

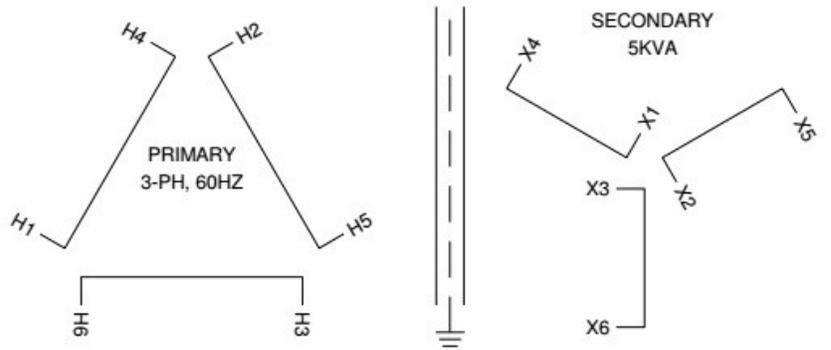

240V DELTA:
INPUT ON LINES H1, H2, H3
JUMP H1-H6, H4-H2, H5-H3

416/240V WYE:
INPUT ON LINES H4, H5, H6
JUMP H1-H2-H3

277V DELTA:
OUTPUT ON LINES X1, X2, X3
JUMP X1-X6, X4-X2, X5-X3

480/277V WYE:
OUTPUT ON X4, X5, X6
JUMP X1-X2-X3

Fig 2.16: Step up transformer.

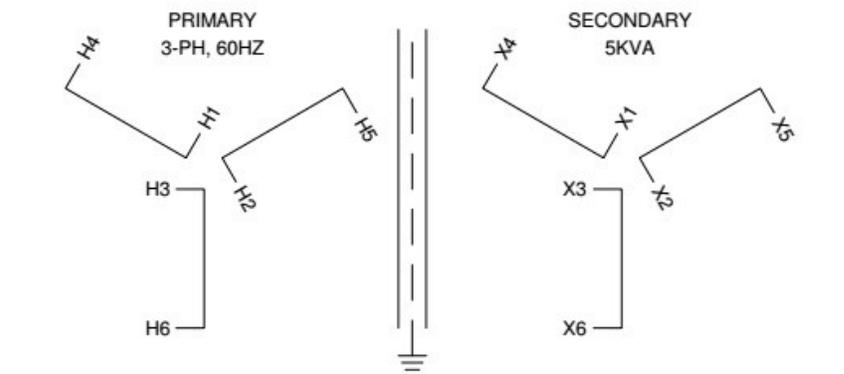

277V DELTA:
INPUT ON LINES H1, H2, H3
JUMP H1-H6, H4-H2, H5-H3

480/277V WYE:
INPUT ON LINES H4, H5, H6
JUMP H1-H2-H3

139V DELTA:
OUTPUT ON LINES X1, X2, X3
JUMP X1-X6, X4-X2, X5-X3

240/139V WYE:
OUTPUT ON X4, X5, X6
JUMP X1-X2-X3

Fig 2.17: Step down transformer.

The system uses a mutual ground for all power and measurement circuit components. This assures an equal reference point in all data reporting and acquisition. The ground end connection boxes are shown below.

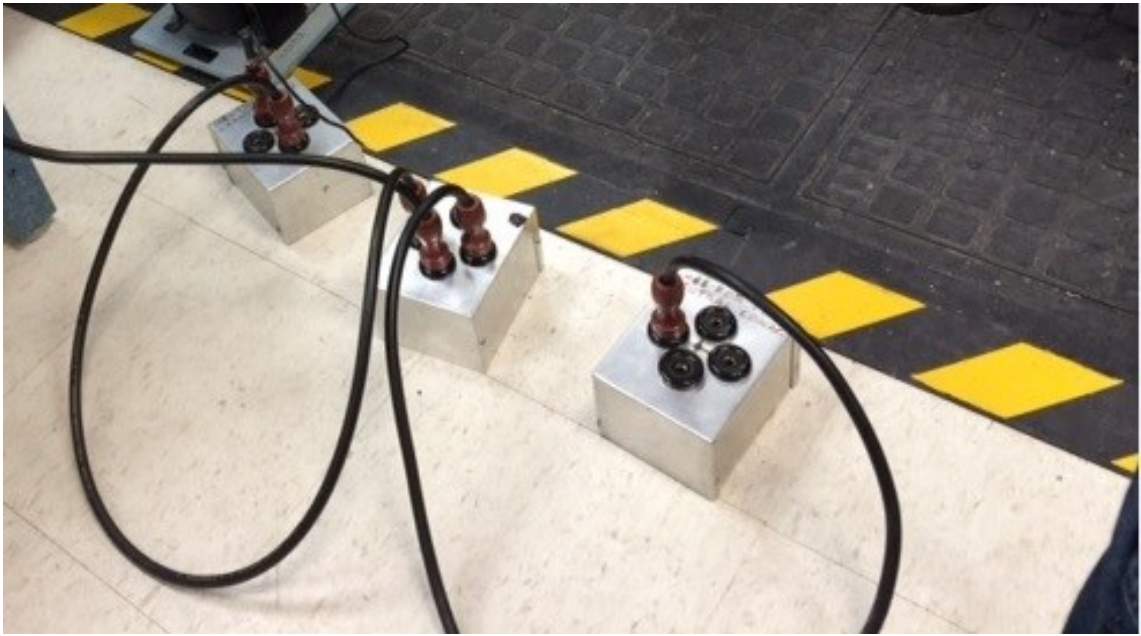

Fig 2.18: Ground connections of the system.

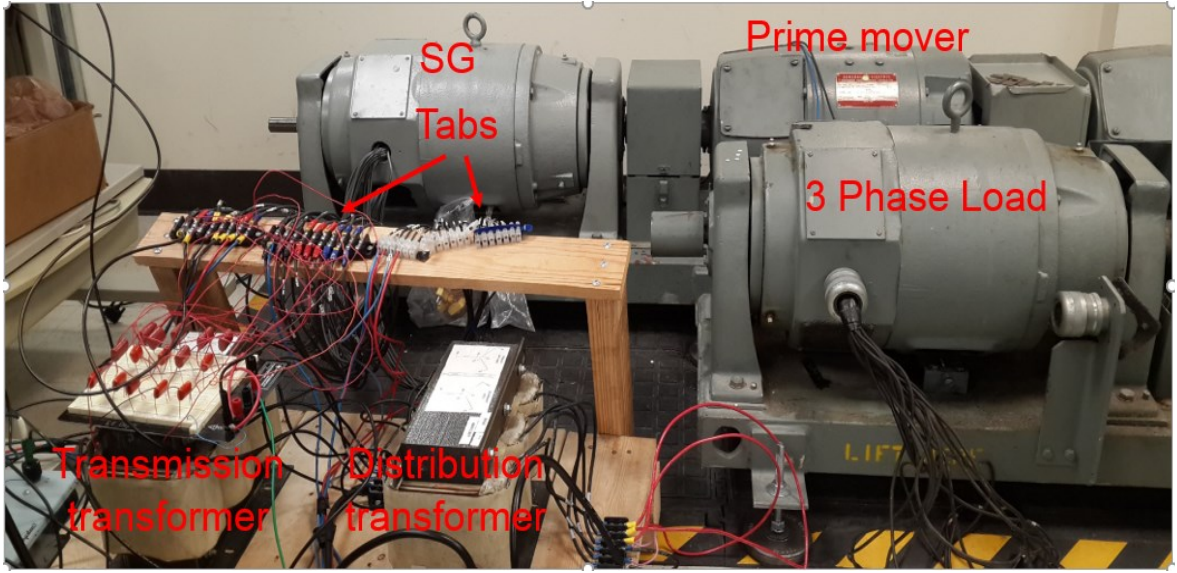

Fig 2.19: Power system components for the lab generating station, adapted from [39].

The neutral grounding transformer is rated at 120/240 V and its low voltage side is connected to the measurement circuit whereas its high voltage side is connected to the generator neutral. This assures a high impedance grounding path since the neutral resistor is four times of the actual resistor. The neutral resistor value is calculated by [33]:

$$R_n = \frac{1}{N^2 2\pi f C} \qquad (2.13)$$

where $R_n$ is the neutral resistor, N is the grounding transformer turns ratio, f is the fundamental frequency and C is the total parasitic capacitors in all three phases. The resulting value of $R_n$ in this system will be 88Ω. The step up transformer is a 240/480 V with Δ-Y connection whereas the step down transformer is rated at 480/240 V with Y-Y connection. The delta connection of the primary of the step up transformer avoids $3^{rd}$ harmonic flow to the grid. The three phase load is a resistive Y connected load with 10Ω resistance at each leg. Taps are available at all three phases windings and in distances of 3, 6, 9, 12, 25, 37.5, 50, 62.5, 75 and 87.5 percent of the total winding length. The fault resistance is chosen to be 50Ω. Fig. 2.20 demonstrates the actual rotor and stator prototype.

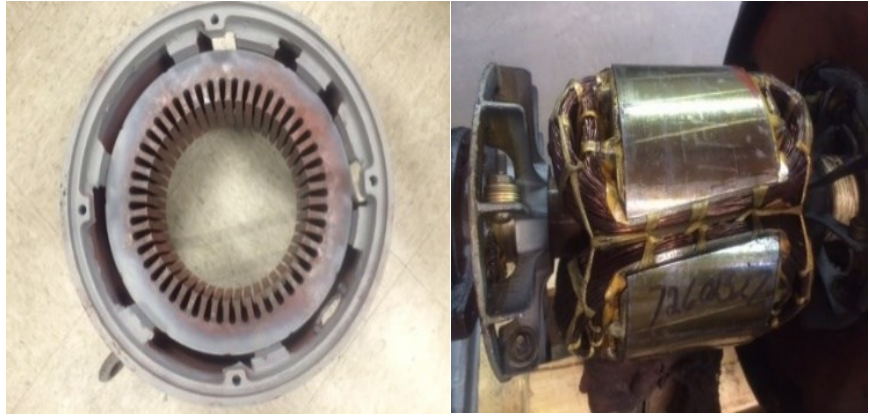

Fig 2.20: SG stator (left) and rotor (right) prototype.

The original synchronous generator stator winding function layout shown below had a very low third harmonic component distribution factor. Phase A, B and C are demonstrated in black, red and blue, respectively. This extremely low third harmonic content in the air gap was not sufficient for third harmonic differential voltage scheme implementation since the expected fault behavior was not emulated as shown in Fig. 2.22 for two types of hard terminal and hard neutral faults. Both terminal and neutral third harmonic voltages reduce in the neutral fault case.

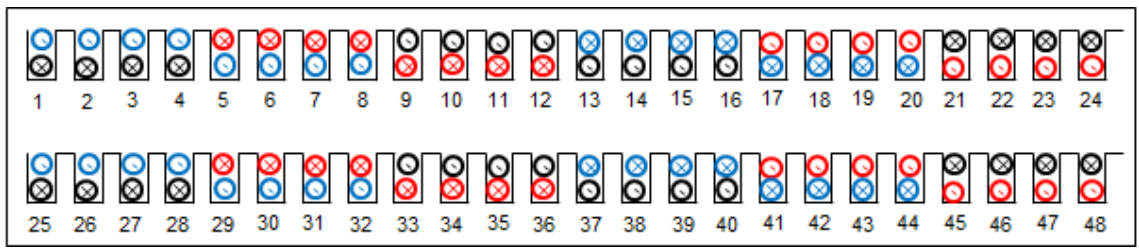

Fig 2.21: SG original stator winding diagram.

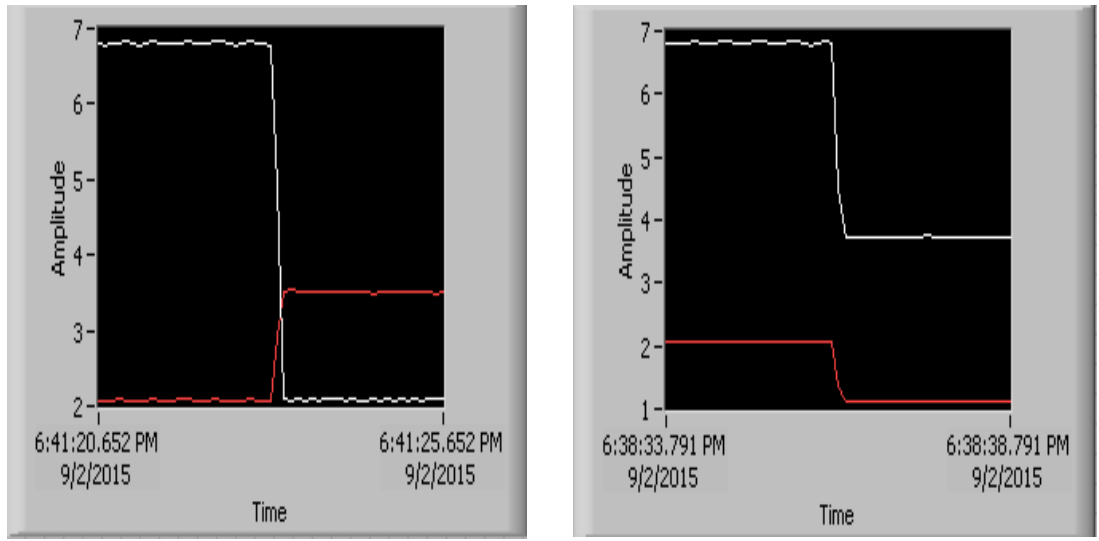

(a)                                         (b)

Fig 2.22: 3$^{rd}$ harmonic voltage behavior during fault at a) phase terminal, and b) neutral.

Therefore, the stator winding configuration was changed to the one shown in Fig. 2.23 which had an increased level of third harmonic distribution factor.

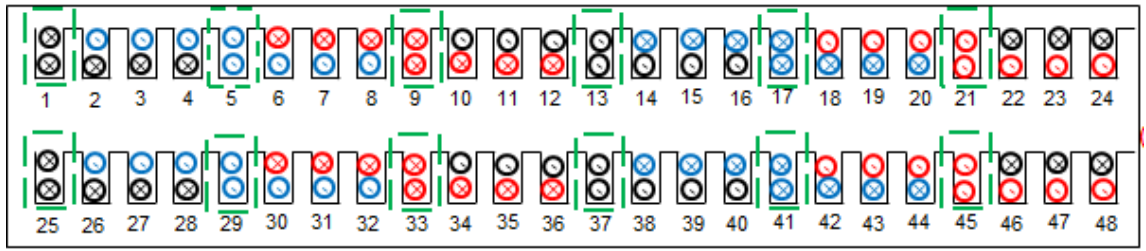

Fig 2.23: SG updated stator winding diagram.

The one slot overlap in each phase layer is highlighted in the figure. The harmonic content of air gap magnetic flux for both of the original and updated stator windings are shown in Fig. 2.24. Almost zero third harmonic voltage is generated in the original winding whereas it is increased to 0.25 after rewinding.

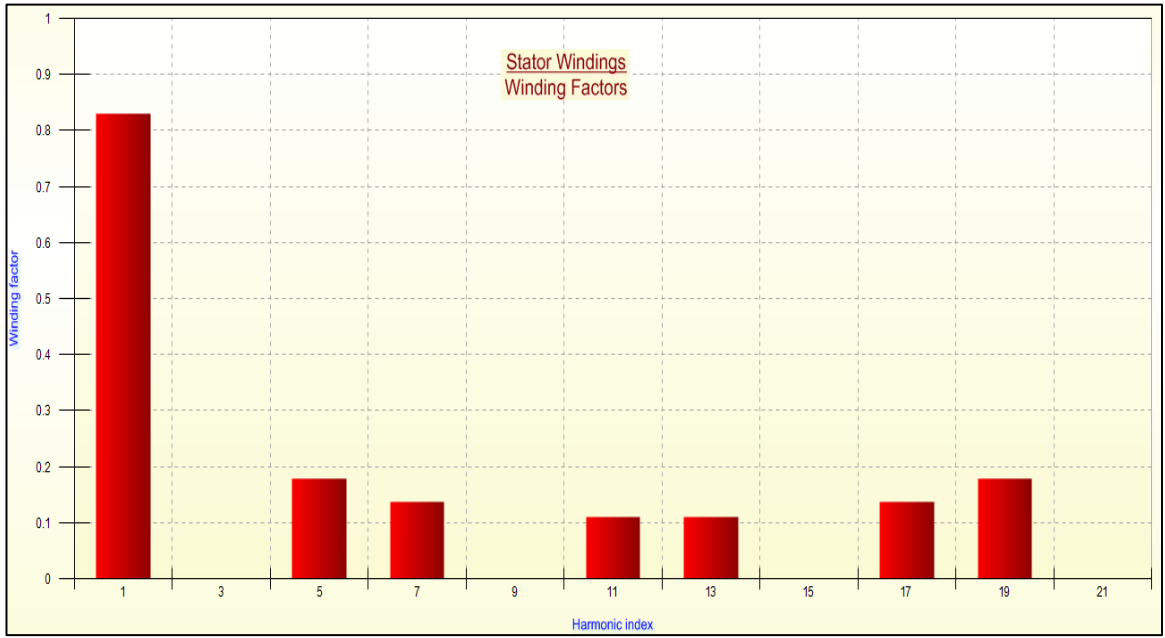

(a)

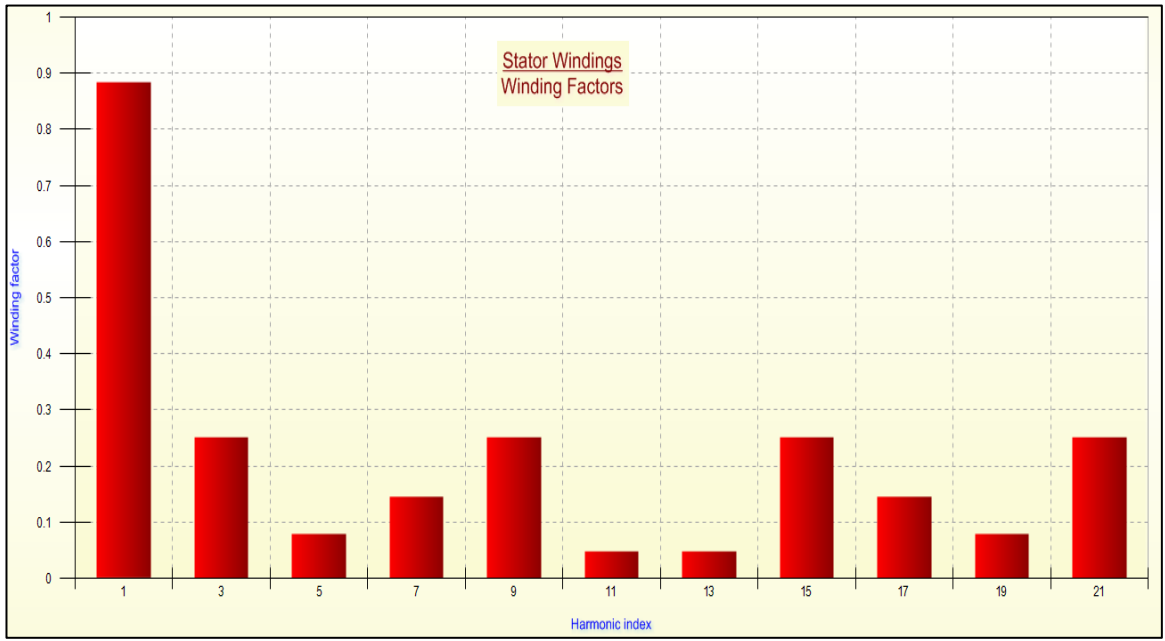

(b)

Fig 2.24: Winding factor FFTs, a) original winding, and b) updated winding.

A dc motor is used as the SG prime mover. A step-up transformer transfers the generated power from the SG to the transmission line where at the end a step-down transformer brings the voltage down for distribution purposes. A synchronous condenser is connected at the output as the load. It can deliver the power factors ranging from 0.8 lagging to 0.8 leading.

• Data acquisition and protection unit: Three phase connected measurement potential transformers (PTs) at the machine terminal and a neutral grounding circuit are the locations where the LabView data acquisition system is connected. Different signals are captured and analyzed in real time to provide any protection algorithm with the required data. This subsystem in the actual setup is shown in Fig. 2.25. Three 120/5V PTs were used to scale down the generator output voltage in order to be interfaced safely with the measurement circuit. A sample acquired waveform by LabView at non-faulted operation of the SG is also shown in Fig. (2.25-b).

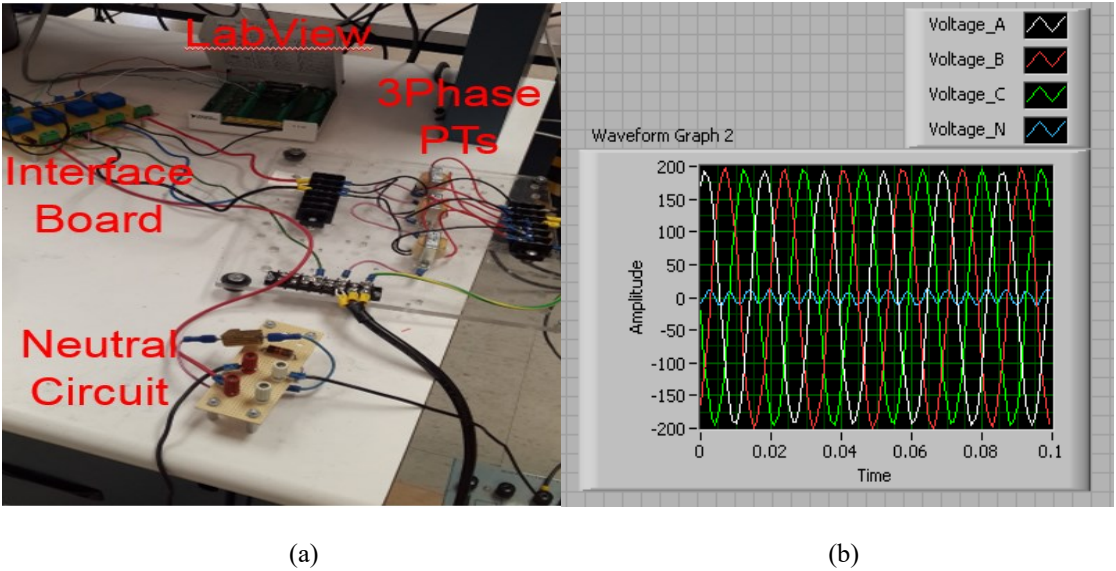

(a)                  (b)

Fig 2.25: Control and protection unit, a) components, and b) sample waveform.

• Fault injector circuit: Several branches connect available stator windings taps to ground via fault resistors. Using switches, the setup is capable of applying various combination of faults on the SG stator windings.

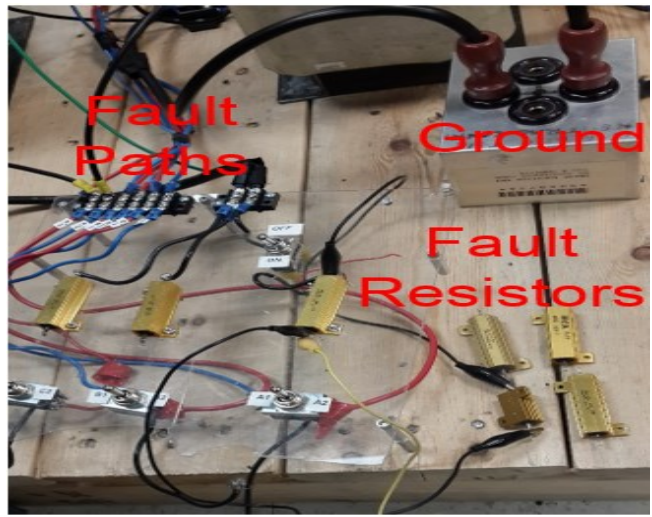

Fig 2.26: Fault injector circuit components.

The original stator windings were pulled out of the slots and rewound to host certain taps at different positions along the winding where the manual faults were applied. A dc-motor is used as the prime mover of the synchronous generator. LabView data acquisition hardware collects required data in discrete form at sampling frequency of 1 kHz. A schematic of created taps along one of the phases' windings is shown below:

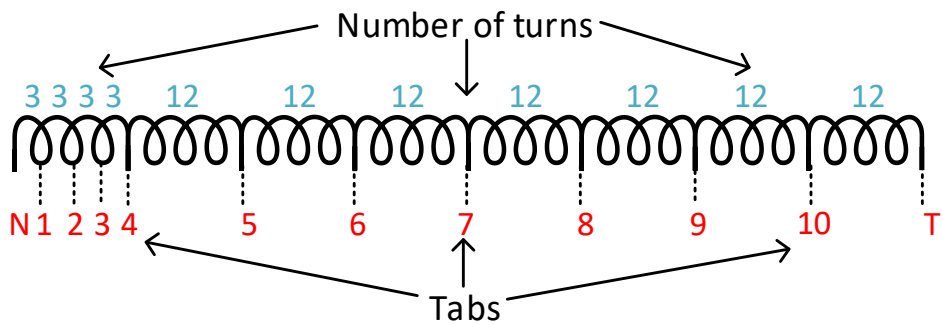

Fig 2.27: Stator windings' taps, adapted from [39].

At the vicinity of neutral, taps are placed every 3 turns whereas towards the terminal end, taps in every 12 turns are used. This is due to higher priority given to the study of faults at or in the vicinity of the neutral point.

By processing acquired data, the primary side third harmonic content of all measured signals are extracted. The unfaulted and faulted data under hard neutral and hard terminal fault (the same as simulation results) were captured and used as inputs to the adaptive 64G2 scheme.

Again, based on trip signals shown in Figs. 2.28 and 2.29, the same responses as in simulation study is observed. The adaptive 64G2 scheme detects both neutral and terminal faults and the sharp crossover at around sample point 12 is not picked up as a fault.

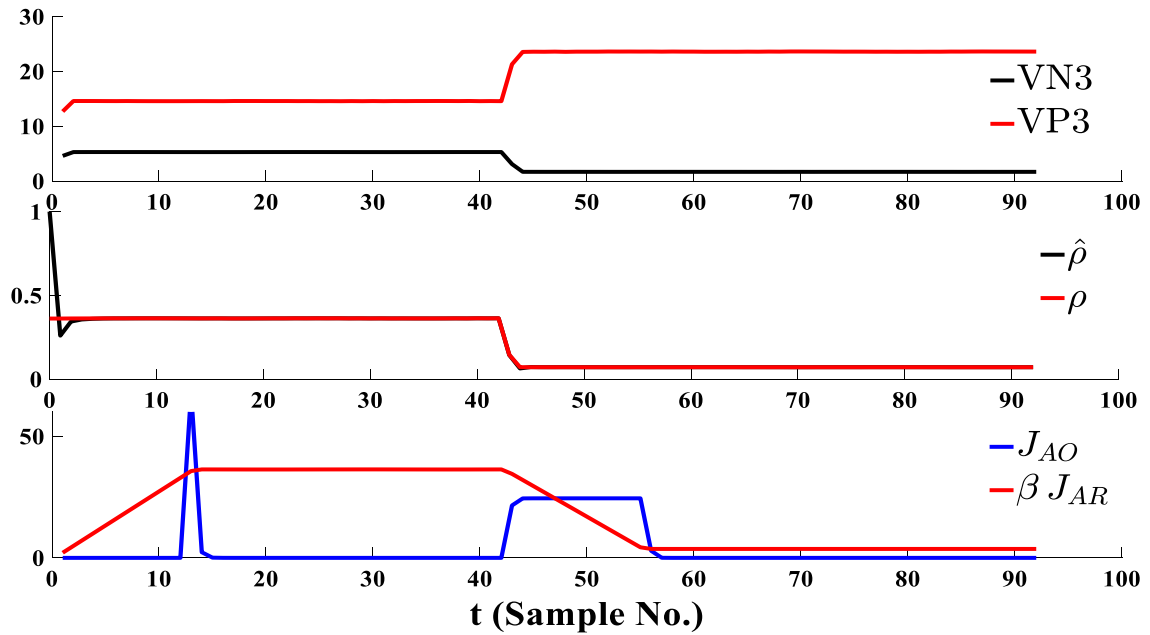

Fig. 2.28: Experimental responses of adaptive 64G2 scheme to hard neutral fault, adapted from [39].

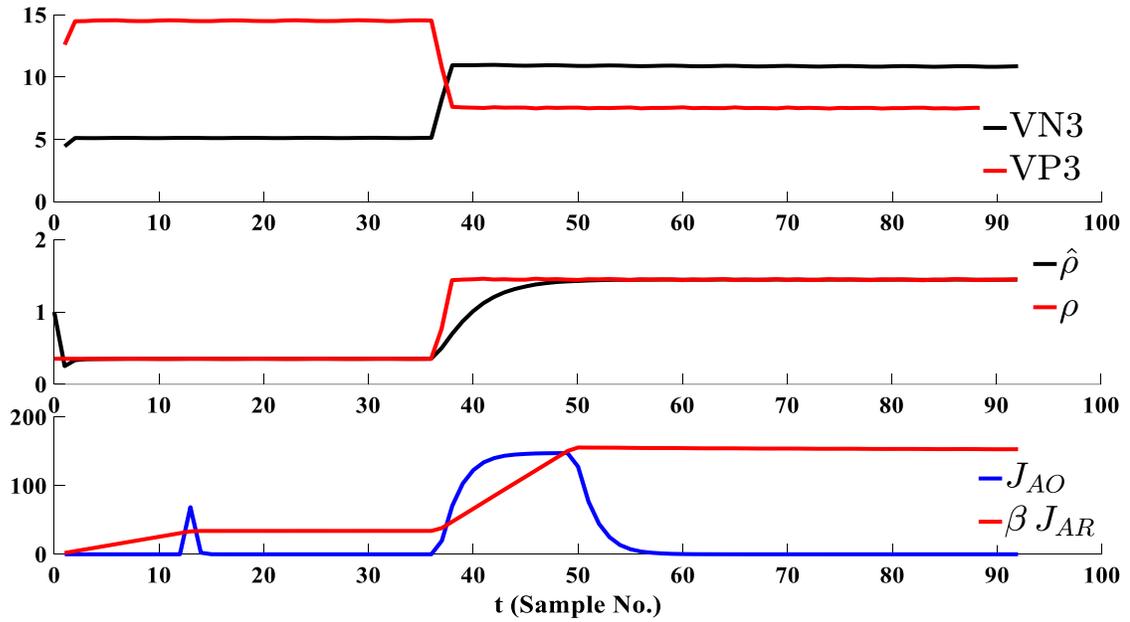

Fig. 2.29: Experimental responses of adaptive 64G2 scheme to hard terminal fault, adapted from [39].

Results according to two extreme taps covering 0% and 100% of the stator winding is tested to verify the accuracy of the adaptive scheme at extremum points. As mentioned earlier, the third harmonic differential voltage scheme has a blind zone near or in the vicinity of the middle of the winding. Several tests at different loads and power factors were done on available taps to find the blind zone of this specific machine. The output of the synchronous generator is connected to a synchronous motor. Different power factors can be achived by variation of the synchronous motor field excitation. A range of 0.8 lag to 0.8 lead is obtainable in this specific system. The synchronous motor's shaft is mechanically coupled to a dc generator whose output is connected to a manually controlled resistive load. Different loading of the synchronous generator is possible by applying different resistive loads.

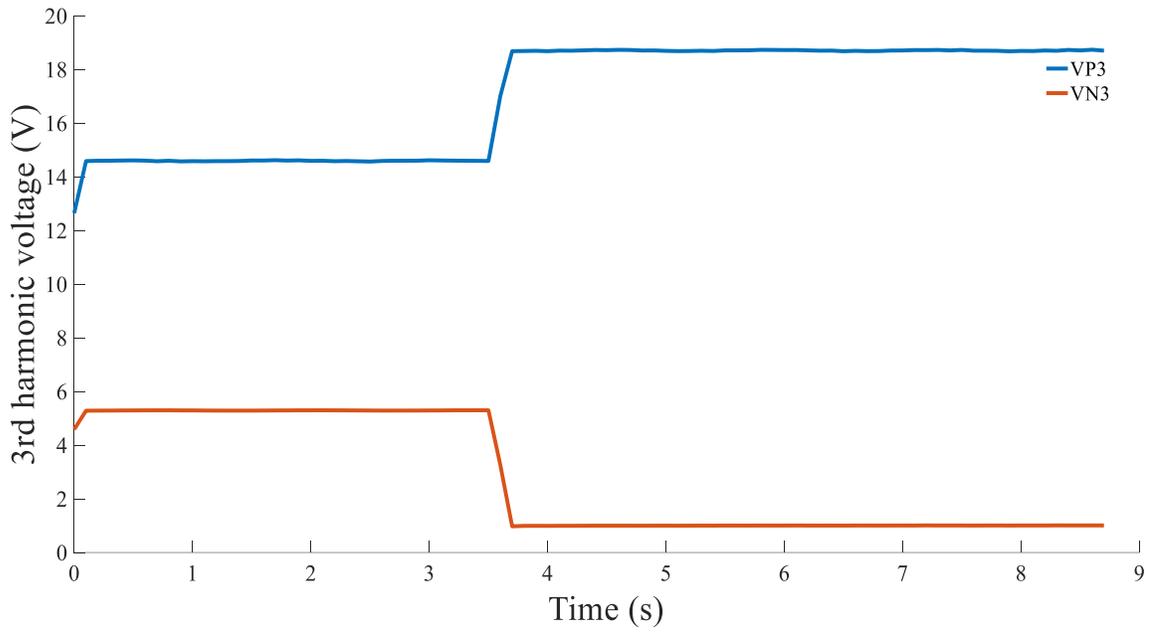

(a)

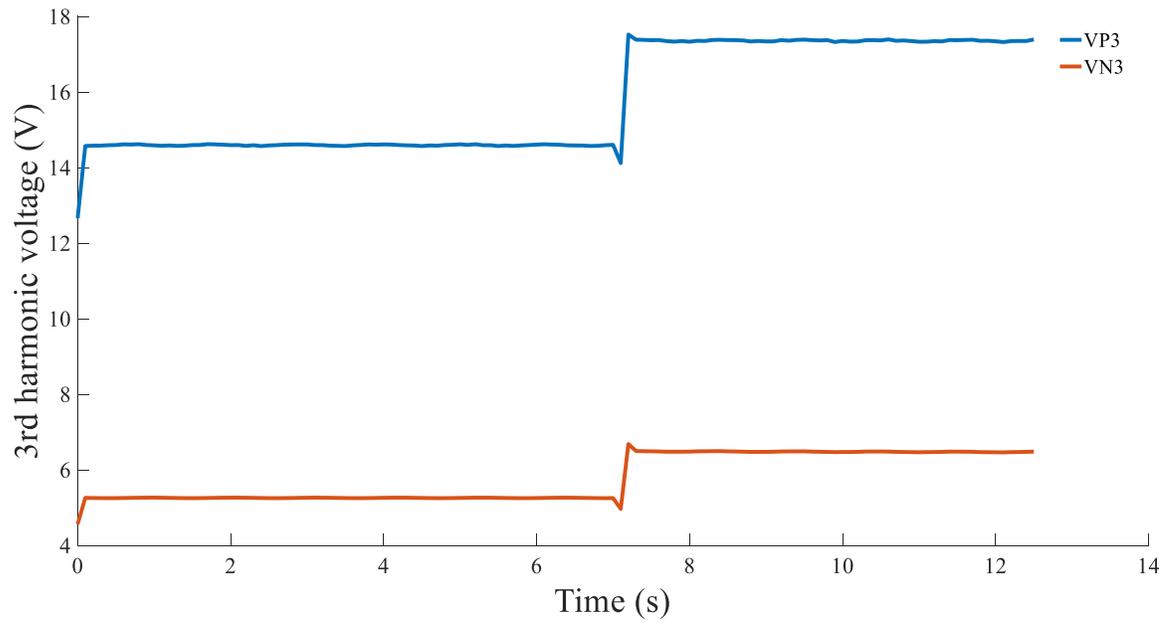

(b)

Fig 2.30: SG fault characteristics at tabs on a) 25%, b) 37%, c) 50%, d) 62% of winding, adapted from [39].

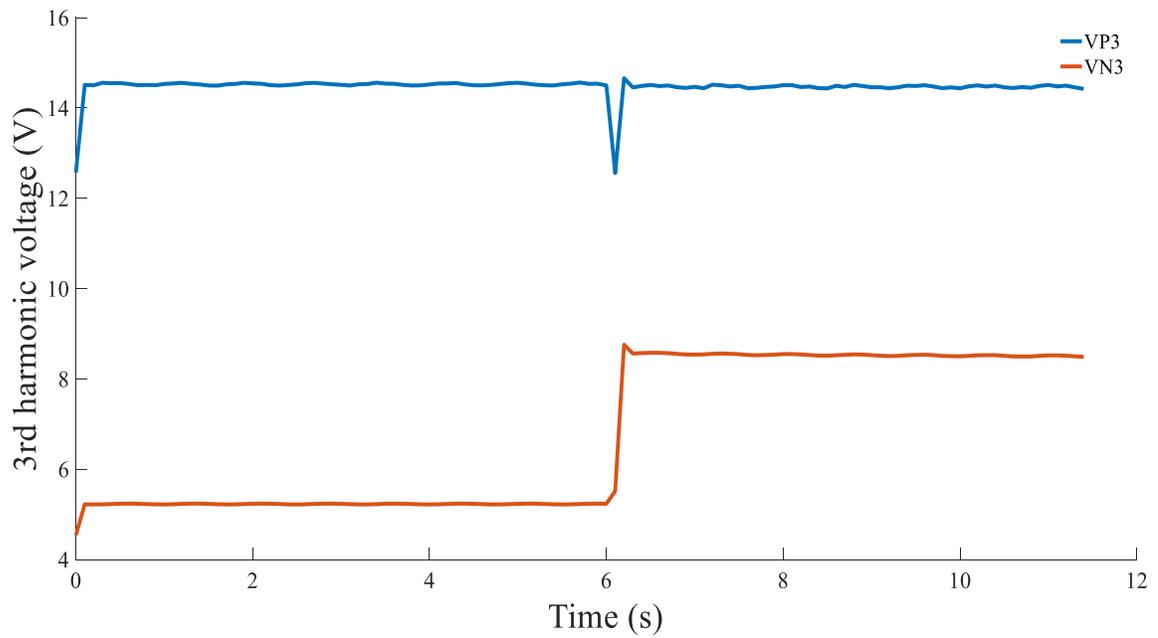

(c)

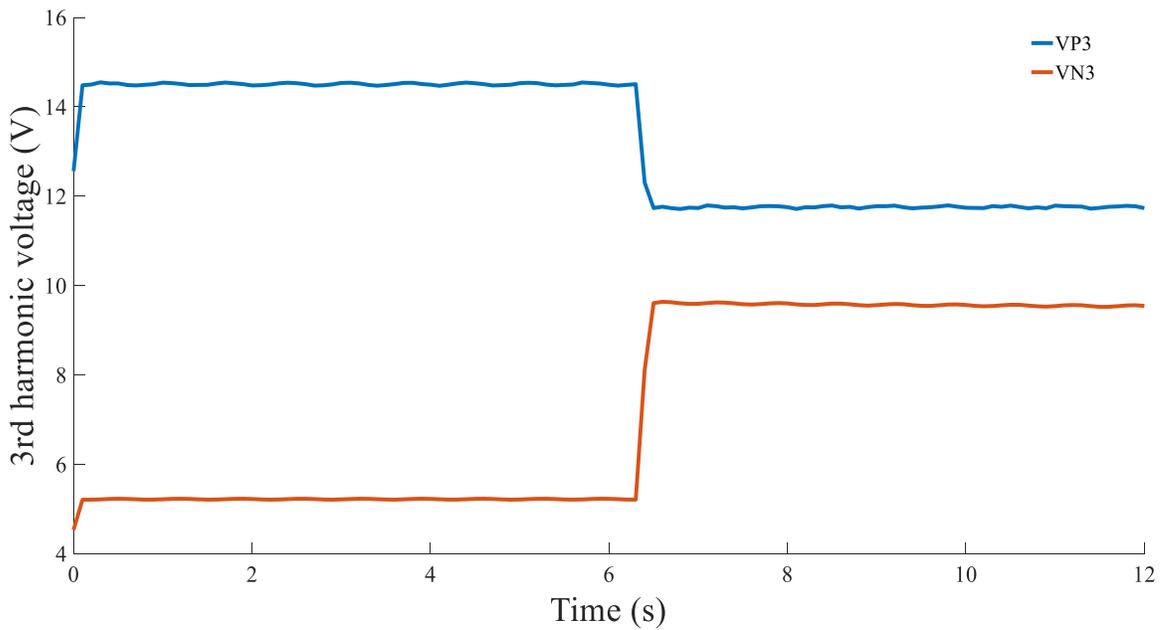

(d)

Fig 2.30 Continued.

Expected fault behaviour at all tabs below 25% and above 62% is seen. Therefore, at least 25-62% is the blind zone in this system. The larger and asymmetric blind zone in the real system compared to the simulation is due to many non-idealities existing in the real system. More resolution of blind zone cannot be achieved due to limitations on available taps.

# 3 RELIABILITY ANALYSIS

## 3.1 Introduction

In a recent paper [30], the industry standard stator ground fault protection based on third-harmonic differential voltages for high impedance grounded synchronous generators (HIGSGs) was scrutinized to show a serious lack of security (the degree of certainty that a relay will not operate incorrectly). Moreover [30] proposed a highly secure fault protection scheme based on the real-time tracking of a residual signal representing the estimated balance between the distributions of the third-harmonic voltage magnitudes at the neutral and terminal ends of HIGSGs. This high level of security was verified using extensive analysis of field collected test data from five utility-size unit-connected HIGSGs with sufficient levels of third-harmonic voltages at both ends of the machine [30]. The key assumption underlying this verification was the textbook characterizations of the neutral and terminal third-harmonic voltages of typical unit-connected HIGSGs during normal and stator ground fault conditions, [14]. Such characterizations are widely used by relay manufacturers to synthesize stator ground protection schemes, [15, 16], [34]. However, it is highly desirable to obtain a convincing verification of the relay scheme's reliability (dependability (the degree of certainty of correct operation in response to system trouble) and security) when it is challenged by real-world stator ground faults, load variations and power system disturbances. While there is no available repertoire of real-world stator ground faults, the importance of considering more realistic fault models has been recognized, [35]. This recognition has led to development of several software-hardware laboratory environments allowing evaluation of relay scheme's reliability before they are incorporated in industrial protective relays, [36, 37]. In the same vein, [38] reported on the construction of a lab-scale generating station which merely tested the dependability of the new A64G2 scheme without any sensitivity and/or security analysis. It is the goal of this chapter to report a completed sensitivity and security study of the proposed A64G2 scheme, [40] and its comparison with the industry standard non-adaptive 64G2 scheme as described in [34] and [39].

## 3.2 Sensitivity

The sensitivity of a stator ground fault protection scheme refers to its fault detection capability when the measured quantities for fault detection are "small". In this sense, the signals used in 64G2 and A64G2 algorithms are needed to be identified. Based on the third-harmonic voltage characteristics in SGs, the amount of variation in stator windings neutral and terminal third-harmonic voltages mainly influence the sensitivity of these schemes. Once an arbitrary point along the winding is connected to the ground via a fault resistor ($R_f$), the induced voltage at that point is applied across $R_f$ resulting in a current through the fault path to the overall system ground. This current will flow back to the winding either through distributed capacitors or the neutral grounding resistor. The voltage variation along the winding due to fault current circulation is the primary indicator in the third-harmonic differential voltage schemes. In order to facilitate studying different stator windings to ground fault scenarios, the original windings had to be rewound to host some taps at several locations along the winding. Through measurement, it was discovered that the total capacitive coupling of the windings to ground is very small (0.008 µF) and cannot emulate the expected third harmonic fault characteristic. Hence, five uniformly distributed capacitances e ach 0.5 µF were added by externally connecting capacitors to the windings taps to improve the third-harmonic fault characteristics. Once the setup was capable of producing reliable data, the improved sensitivity of adaptive scheme over the non-adaptive scheme was tested and compared. Specifically, at full load, A64G2 offers more sensitive coverage than 64G2 at 60, 72 and 84 turns. This is shown in Fig. (3.1-a). Also, at half load operation, it has enhanced sensitivity at 60 turns, see Fig. (3.1-b). Note that the gray region indicates the part of the stator windings which are in the blind zone of the third-harmonic differential voltage for both schemes and 0 and 96 turns corresponds to the neutral and terminal, respectively. The improved sensitivity could be extremely crucial in the sense that low-resistance faults can be detected to avoid any possible damage to the copper insulation layers. Non-detection or delayed detection of such single line to ground faults are known to quickly evolve into multiphase to ground in which case high-impedance neutral grounding does not provide any advantage. In utility size generators the cost of repairs of damaged stator winding can be prohibitively high. The third harmonic voltage level of the synchronous generator depends on both the load level and also the power factor. For example, at light load the third harmonic will be decreased compared to high load operation. Following figures demonstrate this variation at different loads and power factors, [41]:

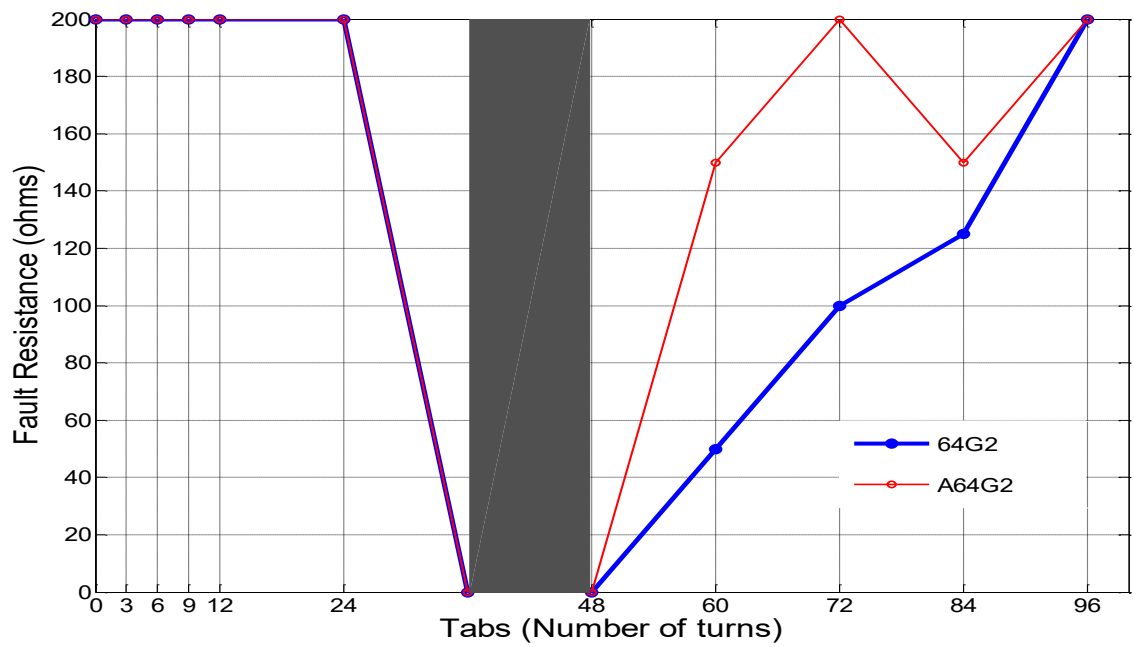

(a)

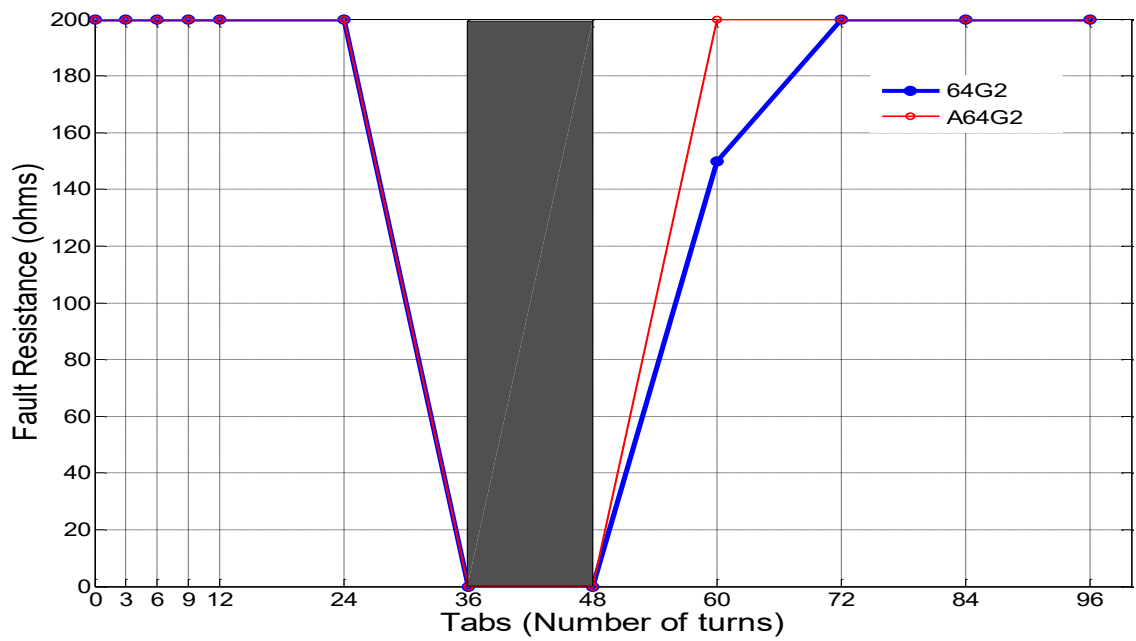

(b)

Fig 3.1: 64G2 and A64G2 sensitivity comparison at a) full-load and b) half-load, adapted from [41].

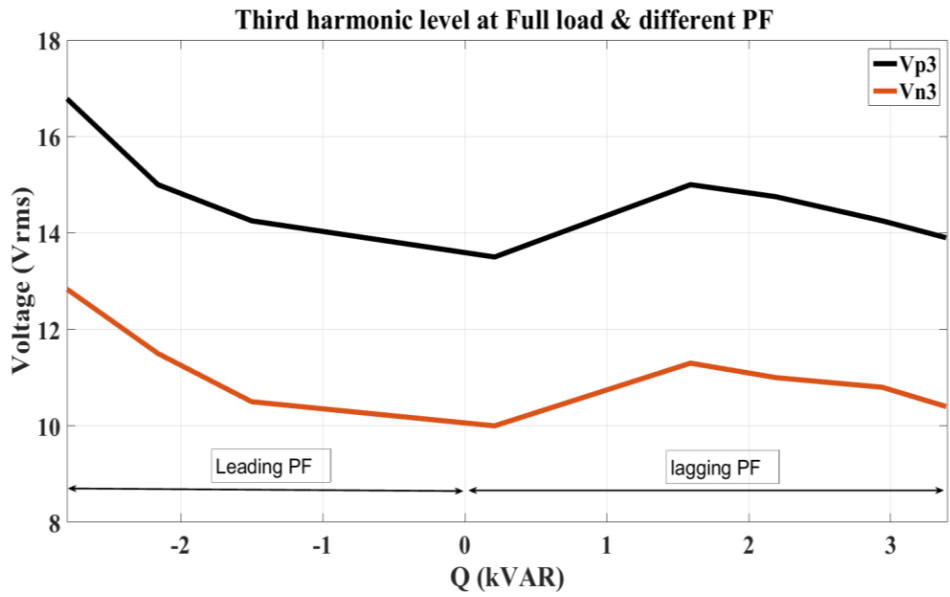

(a)

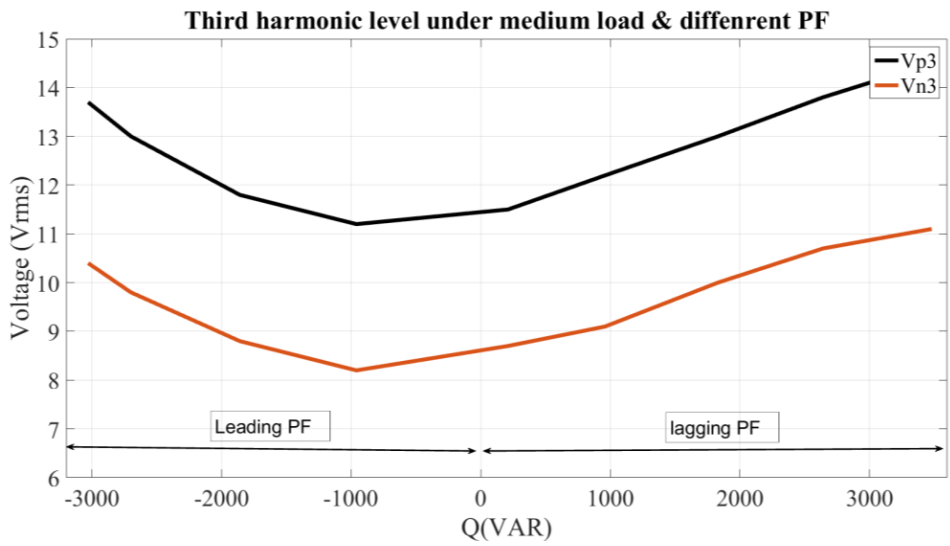

(b)

Fig 3.2: Phase third harmonic voltage variation at a) Full load, b) Half load and c) no load and different power factors.

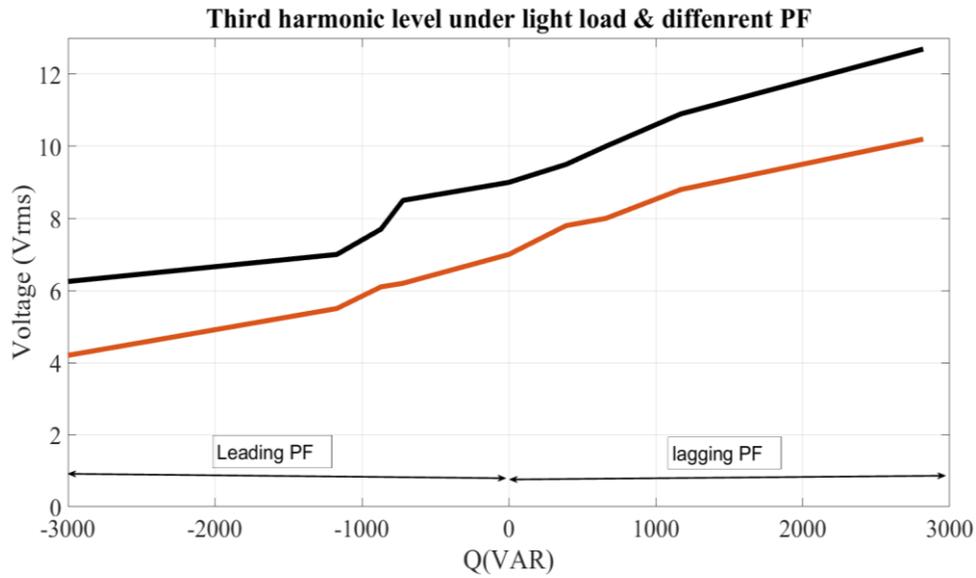

(c)

Fig 3.2 Continued.

## 3.3 Security

Another important feature of a protection scheme is its security, i.e., its robustness against disturbances. It is highly desirable that a protection scheme be secure since otherwise high costs can be incurred as a result of misoperations, i.e. (false trips). Disturbance is attributed to any non-fault phenomenon leading to temporary or permanent variation of third harmonic voltages either at neutral or terminal of the stator windings. Sudden changes in power factor or load levels, out of synch generator breaker closing, phase PTs' performance deterioration and neutral circuit component aging can be mentioned as some possible sources of disturbance. Different disturbances were intentionally created and the adaptive and non-adaptive scheme's responses to them were compared to each other. Here, the neutral circuit disturbances are shown in Fig. 3.4. Clearly, the A64G2 is insensitive to the neutral disturbances at 12.5% change in $V_{N3}$ as shown in Fig. (3.4-a) while the non-adaptive 64G2 nearly trips for this situation. A second scenario with 60% change in $V_{N3}$ as shown in Fig. (3.4-b) results in a misoperation. It should be mentioned that in the 64G2 scheme 64RAT parameter is predetermined using a finite set of data obtained during testing. In particular, for the SG studied in this thesis, eleven test data points were obtained. Fig. 3.3 shows the obtained test data points and the resulting trip boundary lines once 64RAT and $\beta_{64G2}$ are calculated.

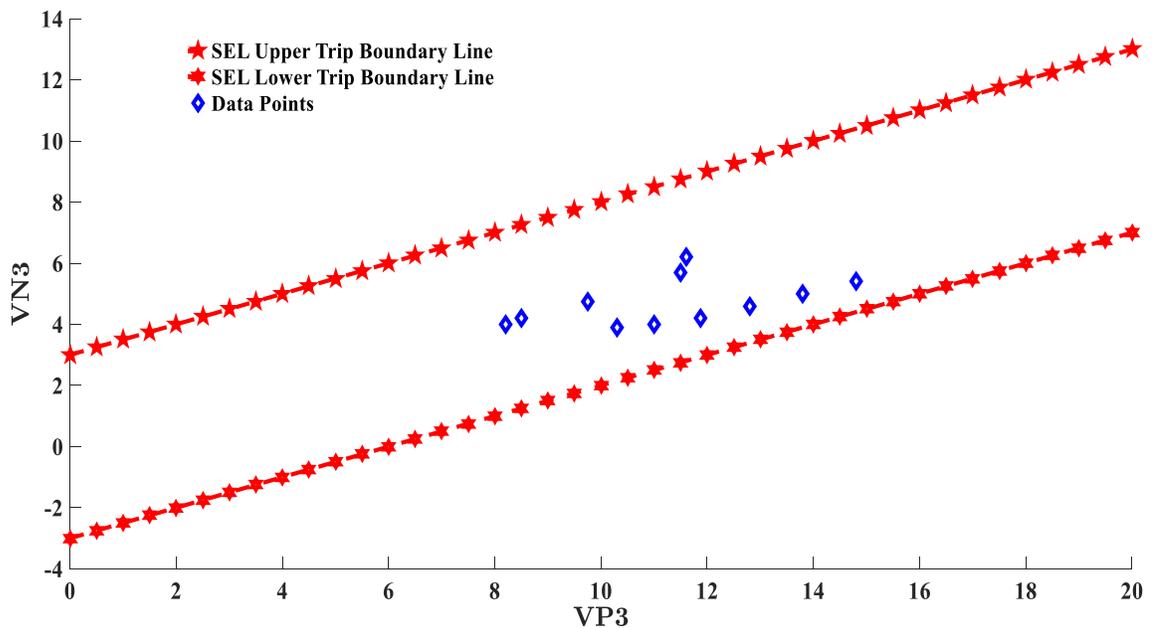

Fig 3.3: 64G2 trip boundary lines and data points, adapted from [41].

Note that data points inside the trip boundary lines indicate healthy operating points while those outside these lines correspond to a stator ground fault. Clearly, due to the limited sets of data points, the real healthy zone of operation can only be approximated. Ideally, a large set of test points are required to attain a "reasonable" estimation of fault-free and faulty operating regions. This points out a weakness in the 64G2 scheme as it may sometimes lead to misoperations discussed in [30]. In contrast to 64G2 scheme, the third harmonic voltage ratio estimate $\hat{\rho}(t)$ in A64G2 is continuously updated during the machine operation. $\beta_{A64G2}$ was set such that the restraint quantities for both adaptive and non-adaptive schemes would be equal before fault. The sharp crossover at the early samples and around sample 70 in Fig. (3.4-b) lasts only for two sample points and will not be detected as a fault by the scheme.

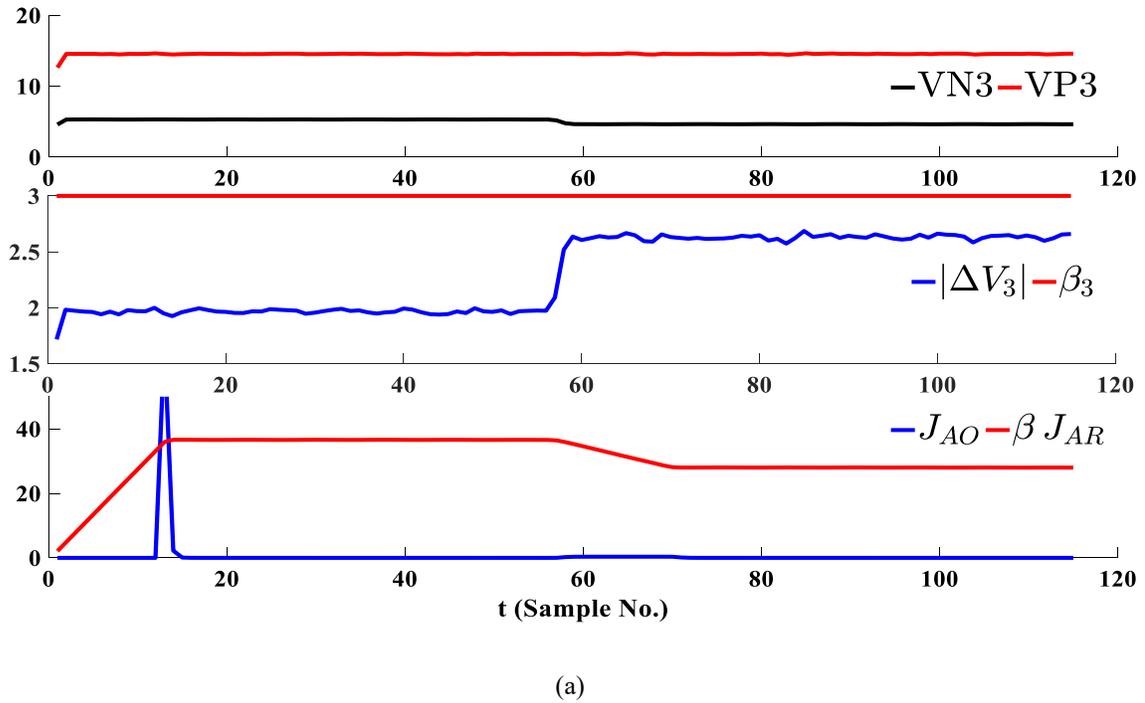

(a)

Fig 3.4: Adaptive and non-adaptive 64G2 security reponses to a) 12.5% and b) 60% neutral circuit disturbance, adapted from [41].

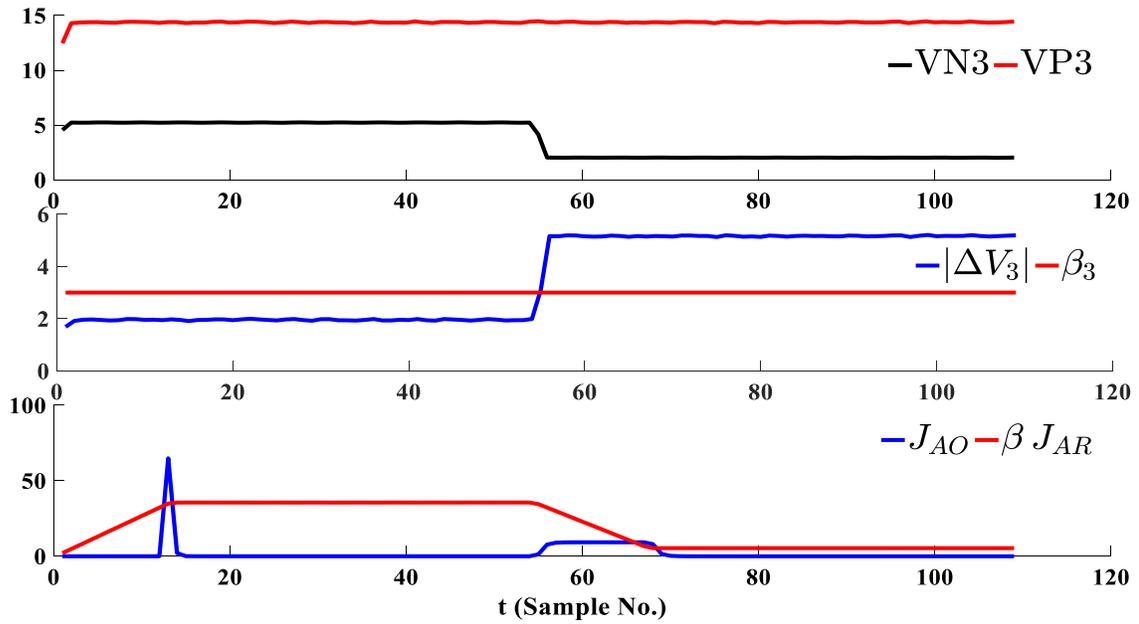

(b)

Fig 3.4 Continued.

Based on above figures, improved security of adaptive scheme to high enough neutral disturbances is observed which is a key finding of this research.

Following figures discuss adaptive and non-adaptive shcemes responses to some probable disturbances in system operation. Fig. 3.5 and 3.6 shows a sudden change in power factor from the most extreme lagging to leading at full and half load. Although the non-adaptive scheme is agitated more but non of them is mislead. It should be mentioned that the third harmonic voltage level is different at full load and half load where it is less at half load. Also, the quality of phase and neutral thrid harmonic voltage variation is different. Fig. 3.7 and 3.8 show a symmetric decrease in load level 4 to 3kW and 5 to 3kW respectively. Apparently when the load decreases, the third harmonic level at both the phase and the neutral decreases. Hence, the net effect results in reduction of operate quantities at both cases. This disturbance would never cause a msioperation. Fig. 3.9 and 3.10 is obtained while the machine is starting to run from standstill condition or when it is speeding down to standstill. At starting, SG eventually reaches at steady state condition which is basically a healthy mode of operation and no trips will be commanded. On the other hand during stopping, since the phase voltages are reduced, the third harmmonic voltages will be also reduced and such as load reduction scenario, no misoperation is happened. Fig. 3.11 correspond to disturbance at phase PTs. It can be deduced that for phase PT disturbances since the voltage on a certain phase is reduced, the same scenario as load reduction or generator stopping is applicable. That is the operate quantity is agitated but is moved away of the restraint quantity so no misoperation risk is expected.

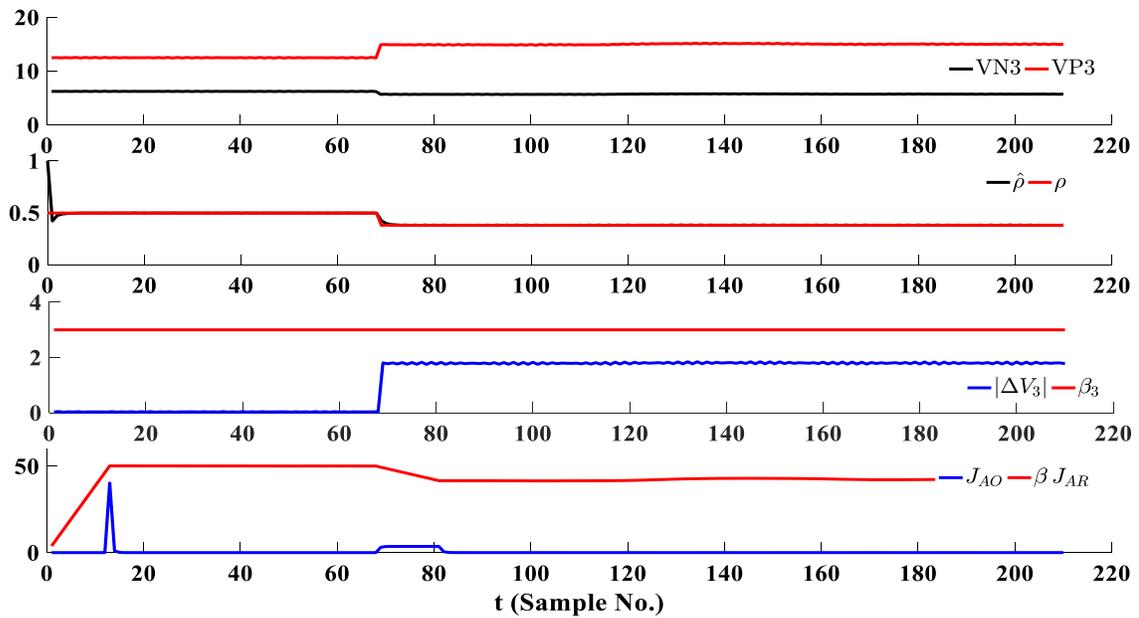

Fig 3.5: Load power factor change from 0.85 lag to lead at full load.

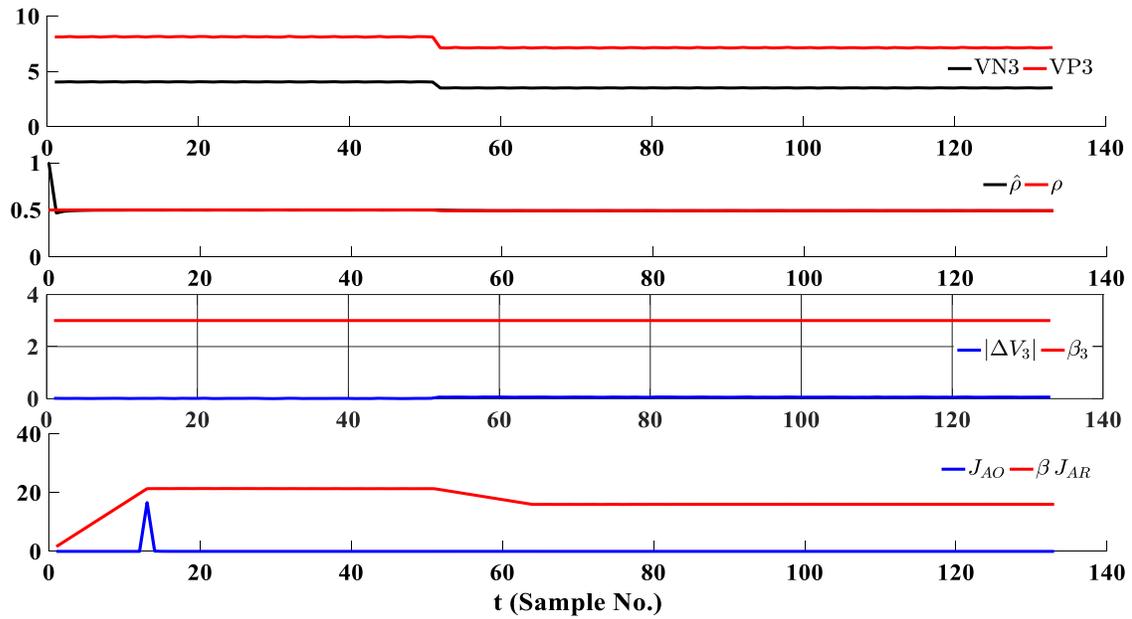

Fig 3.6: Load power factor change from 0.85 lag to lead at no load.

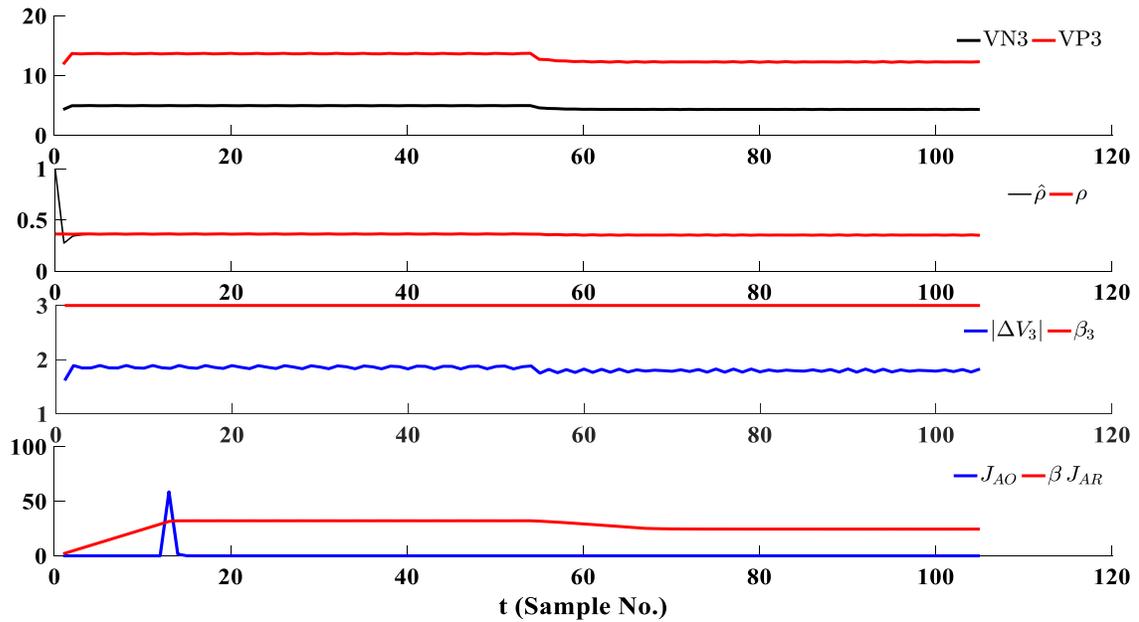

Fig 3.7: Load change from 4kW to 3kW.

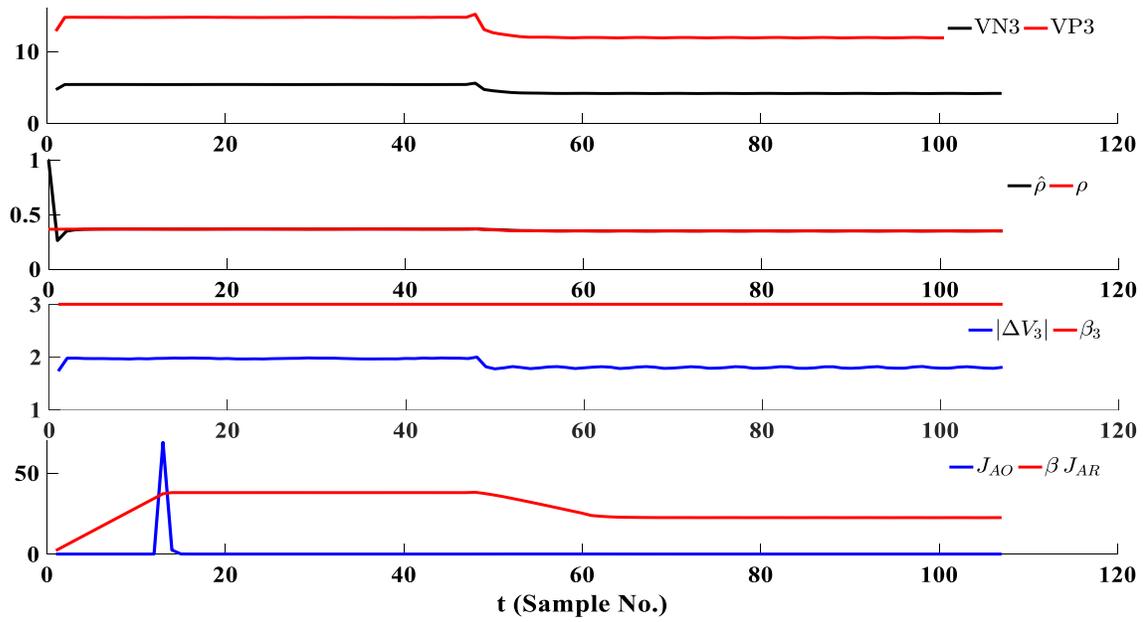

Fig 3.8: Load change from 5kW to 3kW.

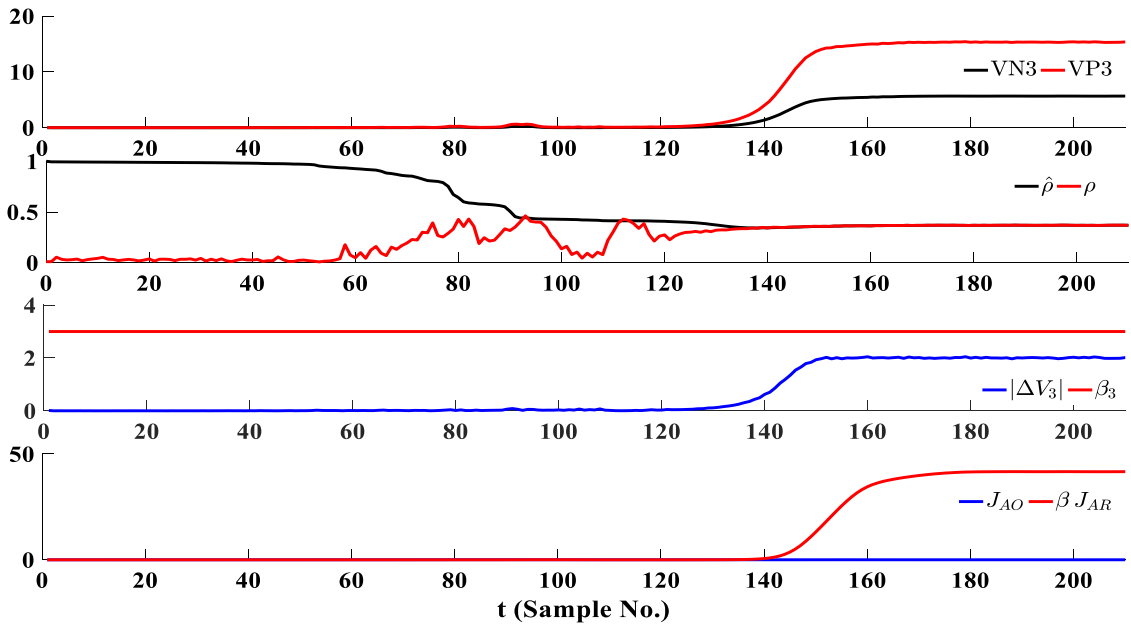

Fig 3.9: Generator starting.

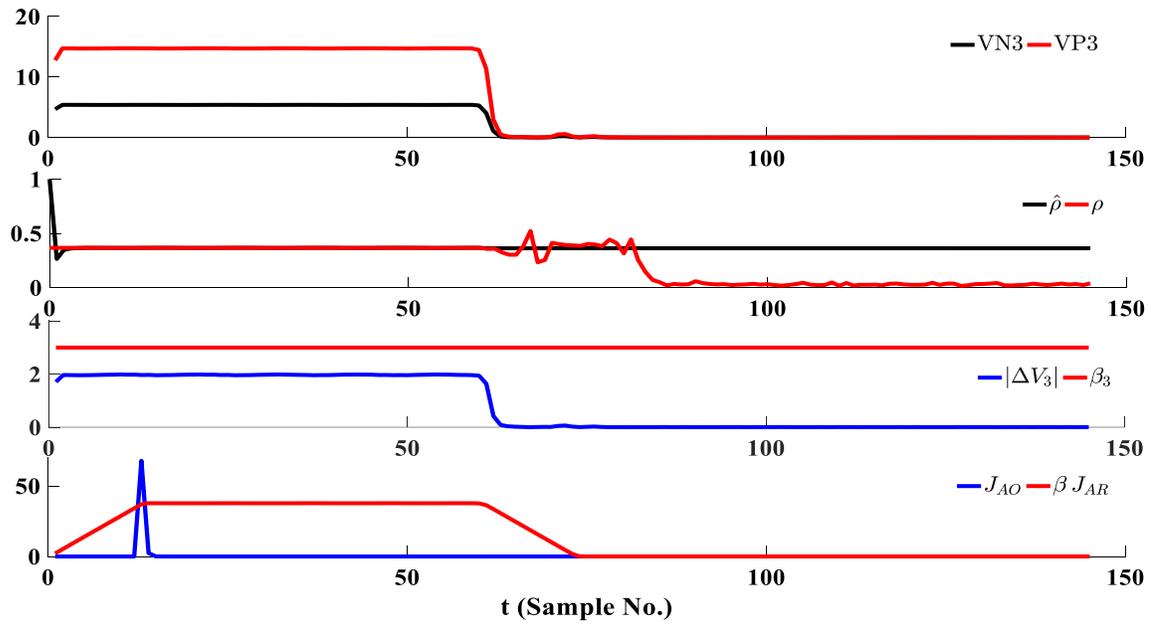

Fig 3.10: Generator stopping.

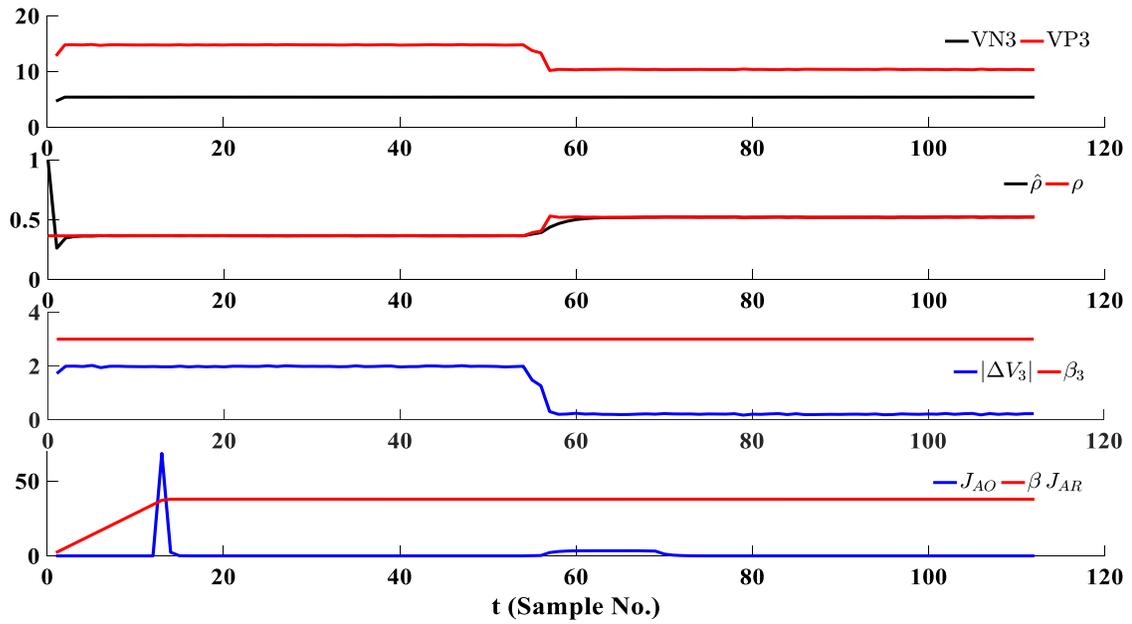

Fig 3.11: Terminal 60% PT disturbance.

# 4 SUB-HARMONIC VOLTAGE INJECTION

The third harmonic differential voltage scheme described in the previous section is commonly used in conjunction with the neutral fundamental overvoltage scheme to protect 100% of the stator windings. In some synchronous machines, the third harmonic voltages at the generator neutral and terminal vary due to generator loading, power factor variation and system disturbances. These variations make the third harmonic schemes insecure resulting in false alarms or misoperations leading to costly loss of production and testing of the generator, [38, 39]. The sub-harmonic voltage injection scheme (64S) is believed to protect 100% of the stator winding, [17 - 20]. However, its application depends heavily on accurate identification of the total capacitance to ground of the generator stator windings, iso-phase bus work and delta-connected windings of the step-up transformer, [43, 44]. Since the operation of the conventional subharmonic injection based protection schemes are arguable in terms of sensitivity, [45-47], some other novel schemes were proposed to resolve the lack of sufficient sensitivity, [19, 48, 49] in which offline testings are required to adopt an accurate estimation model. Therefore, an adaptive scheme based on subharmonic injection is proposed and examined for a fictitious unit with pre-assumed system parameters in [20]. This novel scheme estimates the values of insulation parameters such as the total equivalent insulation resistance of the machine, equivalent stator windings to ground capacitance and the time constants based on Kalman adaptive filtering theory, [50]. The experimental verification of the adaptive scheme proposed in [20] for high impedance grounded synchronous generators is also performed. To this end, a laboratory scale synchronous generator setup is designed and built to closely emulate the behavior of industrial synchronous generators such that subharmonic injection facilities could also be equipped on it. The data from fault studies has so far established the dependability of the adaptive scheme, i.e., correct operation due to faults. The outstanding advantages of the A64S scheme compared to the non-adaptive schemes are also explained through the chapter. This chapter briefly reviews the operational theory in subharmonic injection methods. Then, the adaptive scheme is described in detail. The experimental results for different fault cases and scenarios are discussed in the last section.

## 4.1 Subharmonic Injection Method Description and Parameter Selection

In order to simplify the analysis of the system, the equivalent circuit including the key parameters contributing in the detection scheme shown in Fig. 4.1 are derived. The parameters are also represented in Table 4.1.

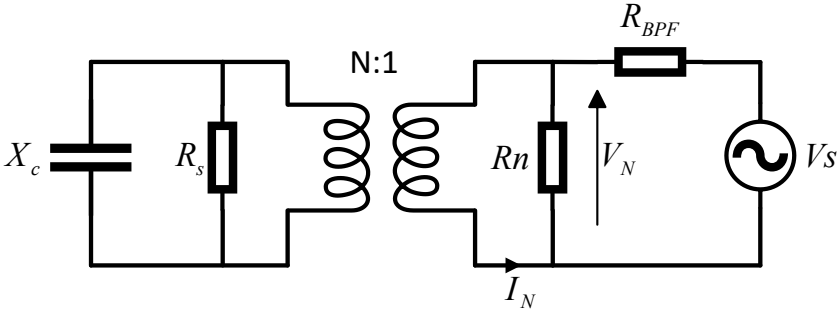

Fig. 4.1: Overal system equivalent circuit.

Table 4.1: Equivalent circuit parameters

| Parameter | Description | Value |
|---|---|---|
| Xc | Total machine and connected apparatus capacitance | To be estimated. |
| Rs | Total insulation resistance | To be estimated. |
| N | Neutral transformer turns ratio | 2 |
| Rn | Neutral resistor | 250 Ω |
| Rbpf | Band pass filter series resistance | 8 Ω |
| Vs | Subharmonic voltage source | 25 Vrms |

The neutral grounding resistor which assures the high impedance grounding of the system is chosen to be 250Ω in order to build up a sufficiently high neutral current. A subharmonic voltage generator injects a 20Hz voltage across the secondary of the neutral grounding transformer. This voltage induces a 20Hz current through the neutral circuit and consequently the stator windings. In conventional overcurrent based relays, usually the neutral current (In) is passed through a current transformer and measured by the protective relay to decide if it is higher than a pre specified threshold. In this paper, 'In' is directly measured and fed into the adaptive scheme. In order to further simplify the analysis of the system responses to the injected 20Hz voltage, the equivalent circuit in Fig. 4.1 is depicted in the form of Fig. 4.2 with the machine parameters transferred to the secondary of the neutral grounding transformer.

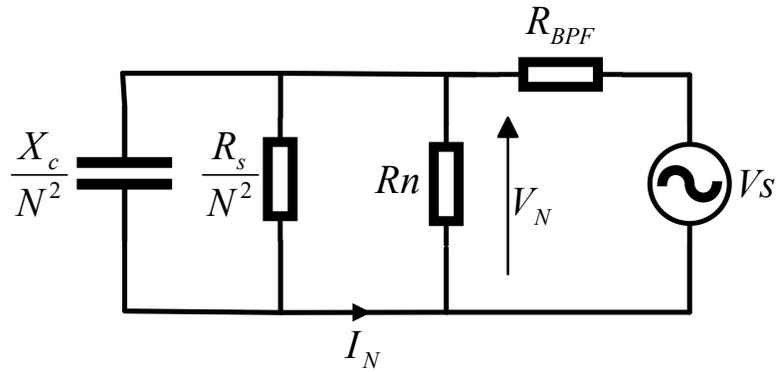

Fig. 4.2: Final simplified equivalent circuit.

The only element in this circuit which should be designed is $R_n$ whereas all others are predetermined based on the experimental prototype.

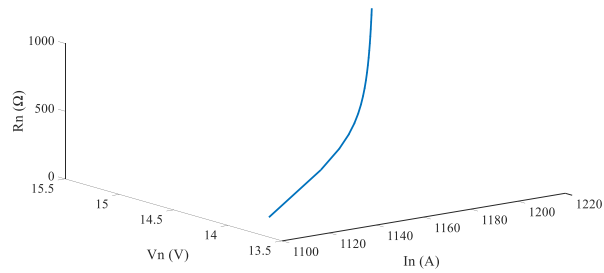

(a)

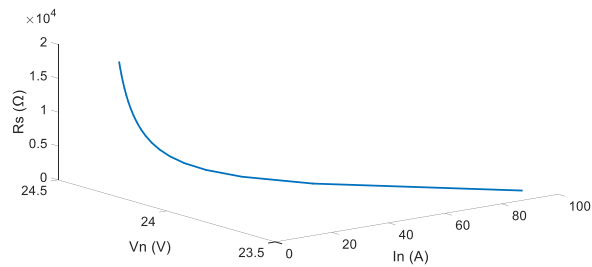

(b)

Fig. 4.3: Neutral circuit voltage and current versus a) neutral resistor, b) insulation resistance, and c) coupling capacitance.

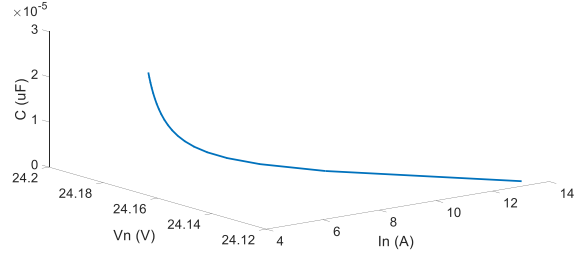

(c)

Fig. 4.3 Continued.

Moreover, since $R_s$ and $C_0$ are not yet recognized for the current system, the variation of $I_n$ with respect to the three parameters ($R_s$, $C_0$ and $R_N$) is represented in Fig. 4.3. Solving the circuit network in Fig. 4.2 yields the transfer function from the subharmonic voltage source ($V_s$) to the neutral current and voltage ($I_n$, $V_n$).

$$\begin{bmatrix} I_N \\ V_N \end{bmatrix} = \begin{bmatrix} H_1(s) \\ H_2(s) \end{bmatrix} V_s(s) \qquad (4.1)$$

where,

$$H_1(s) = K_1 \frac{1+\tau_0 s}{1+\alpha \tau_0 s} \qquad (4.2)$$

$$H_2(s) = \frac{K_2}{1+\alpha \tau_0 s} \qquad (4.3)$$

$$\tau_0(s) = R_s C_0 \qquad (4.4)$$

$$\alpha = \frac{N^2 R_N R_{BPF}}{N^2 R_N R_{BPF} + R_s (R_N + R_{BPF})} \qquad (4.5)$$

and,

$$K_1 = N^2 \left( \frac{R_N}{N^2 R_N R_{BPF} + R_s (R_N + R_{BPF})} \right) \quad (4.6)$$

$$K_2 = \left( \frac{R_N R_s}{N^2 R_N R_{BPF} + R_s (R_N + R_{BPF})} \right) \quad (4.7)$$

**4.2 Adaptive 64S Scheme Theory**

In the proposed adaptive scheme the key identification parameters such as insulation resistance, stator windings to ground capacitance and time constants are estimated continuously to track any change in their values due to real faults or any other sudden disturbance which might cause them to be changed. In contrast, a single calculation is used in conventional subharmonic injection schemes for all operating modes.

4.2.1 System state space model and the proposed KAF

If we consider the transfer function between the neutral current and voltage based on (4.1), the following equation can be obtained:

$$H_3(s) = \frac{V_N(s)}{I_n(s)} = H_2(s) H_1^{-1}(s) = \frac{K_3}{1 + \tau_0(s)} \quad (4.8)$$

where,

$$K_3 = K_2 K_1^{-1} = \frac{R_s}{N^2} \quad (4.9)$$

The above transfer function in continuous-time model should be converted to the discrete-time model using Tustin bilinear mapping:

$$s = \frac{2}{T}\left(\frac{z-1}{z+1}\right) \quad (4.10)$$

where T>0 is the sampling period and 'z' is the z-transform complex variable. Now, the discrete-time transfer function can be represented as:

$$\bar{H}_3(z) = K_d \frac{z+1}{z+a_0} \tag{4.11}$$

where,

$$K_d = \frac{K_3 T}{T + 2\tau_0} \tag{4.12}$$

$$a_0 = \frac{T - 2\tau_0}{T + 2\tau_0} \tag{4.13}$$

Moreover, the corresponding difference equation is given by:

$$V_N(t) = -a_0 V_N(t-1) + K_d u(t) \tag{4.14}$$

with

$$u(t) = I_N(t-1) + I_N(t) \tag{4.15}$$

It should be mentioned that the coefficients of this difference equation are uniquely represented by the original model parameters $R_s$ and $\tau_0$.

To apply the KAF algorithm, we consider the stochastic time-varying parameter model:

$$\theta(t+1) = \theta(t) + \upsilon(t), \quad \theta(0) = \theta_0 \tag{4.16}$$

$$V_N(t) = \phi^T(t)\theta(t) + e_1(t) \tag{4.17}$$

where,

$$\theta(t) = (a_0, K_d)^T \tag{4.18}$$

$$\phi^T(t) = (-V_N(t-1), u(t)) \tag{4.19}$$

We assume $\theta_0$ is a Gaussian random parameter vector with mean $\hat{\theta}_0$ and covariance $\Pi_0$. Moreover, v(t), e₁(t) are zero-mean white noise processes with covariances:

$$E\{v(t)v^T(i)\} = \sigma_v^2 I_2 \delta_{t,i} \quad (4.20)$$

$$E\{e_1(t)e_1(i)\} = \sigma_{e1}^2 \delta_{t,i} \quad (4.21)$$

where, $\delta_{i,j}$ is the Kronecker delta which is equal to 1 if integers i and j are equal, and is zero otherwise. In addition, we assume that $\Theta_0$, v(t) and e₁(t) are uncorrelated.

The justification for the model (4.16) is as follows. Since $\Theta$ represents the unfaulted stator parameters $R_s$ and $\tau_0$, for short sampling periods, its dynamics are fixed, i.e., $\Theta(t+1) = \Theta(t)$. However, since these parameters drift slowly with temperature and time, the process noise v(t) is added to account for these subtle changes. As for the output equation in (4.16), it directly follows from the difference equation (4.14) which describes the interdependence between the imperfectly measured subharmonic neutral grounding circuit signals $V_N$ and $I_N$.

The minimum-variance estimate of $\Theta$ modeled by (4.16) is given by the well-known Kalman filter, [51]. Specifically, for t>1, the KAF algorithm [50] is given by:

$$\hat{\theta}(t) = \hat{\theta}(t-1) + K(t)\nu(t), \quad \hat{\theta}(0) = \hat{\theta}_0 \quad (4.22)$$

$$\nu(t) = V_N(t) - \phi^T(t)\hat{\theta}(t-1) \quad (4.23)$$

$$K(t) = \frac{1}{\sigma_{e1}^2} P(t)\phi(t) \quad (4.24)$$

$$P(t) = \frac{P(t-1)\sigma_{e1}^2}{\sigma_{e1}^2 + \phi^T(t)P(t-1)\phi(t)} + \sigma_v^2 I_2, \quad P(0) = \Pi_0 \quad (4.25)$$

where, $\hat{\theta}(t)$ is the minimum-variance estimate of $\theta(t)$ with $P(t)$ as its error covariance matrix.

Once the parameter $\Theta$ is estimated by:

$$\hat{\theta}(t) = \left[\hat{a}_0(t), \hat{K}_d(t)\right]^T \tag{4.26}$$

the original model parameter estimates can be extracted using:

$$\hat{\tau}_0(t) = \frac{T}{2}\psi\left(F\left[\frac{1-\hat{a}_0(t)}{1+\hat{a}_0(t)}\right]\right) \tag{4.27}$$

$$\hat{R}_s(t) = \frac{N^2}{nT}\psi\left(F\left[(T+2\hat{\tau}_0(t))\hat{K}_d(t)\right]\right) \tag{4.28}$$

Here, 'F' is a low-pass filtering operation with its discrete-time transfer function realization given by:

$$F(z) = \frac{1-e^{-\gamma T}}{z-e^{-\gamma T}}, \quad \gamma > 0 \tag{4.29}$$

and $\psi$ is a saturation operation defined by:

$$\psi(x(t)) = \begin{cases} 0, x(t) < 0. \\ x(t), x(t) \geq 0. \end{cases} \tag{4.30}$$

Note that $F$ is merely used for signal smoothing while $\psi$ is used to discount any negative parameter estimates during initial estimation stages.

Using the original parameter estimations in (4.27) and (4.28), the system coupling capacitance to ground '$C_0$' can be also estimated. As pointed out in [44], $C_0$ is likely to vary with the deformation of the stator windings and voltage changes in the bulk electric system.

The parameter $C_0$ is usually measured as in [44] whereas it is estimated in this approach. No fixed parameter settings for $R_{BPF}$ and $R_N$ is required in the adaptive scheme which is an advantage since these parameters might change with temperature and time and could affect $C_0$ measurement adversely.

It should also be stated that $C_0$ cannot be simply estimated by division of $\hat{\tau}_0(t)$ and $\hat{R}_s(t)$ based on (4.4). To obtain a numerically stable calculation of $C_0$, the following relationship should be assumed:

$$\hat{\tau}_0(t) = \hat{R}_s(t)C_0(t) + e_2(t) \tag{4.31}$$

where $e_2(t)$ is an added white-noise process accounting for the estimation error. Viewing (4.31) as an output equation, we define the parameter $\theta_3 = C_0$, and consider the scalar parameter model:

$$\theta_3(t+1) = \theta_3(t) + \omega(t), \quad \theta_3(0) = \theta_{3,0} \tag{4.32}$$

$$\hat{\tau}_0(t) = \hat{R}_s(t)\theta_3(t) + e_2(t) \tag{4.33}$$

Here, we assume $\theta_{3,0}$ is a scalar Gaussian random parameter with mean $\hat{\theta}_{3,0}$ and variance $\Phi_0$. Moreover, $\omega(t)$, $e_2(t)$ are zero-mean white noise processes with covariances:

$$E\{\omega(t)\omega(i)\} = \sigma_\omega^2 \delta_{t,i} \tag{4.34}$$

$$E\{e_2(t)e_2(i)\} = \sigma_{e2}^2 \delta_{t,i} \tag{4.35}$$

and $\theta_{3,0}$, $\omega(t)$ and $e_2(t)$ are uncorrelated. The KAF estimation algorithm for $C_0$ is given by:

$$\hat{\theta}_3(t) = \hat{\theta}_3(t-1) + K(t)\xi(t), \quad \hat{\theta}_3(0) = \hat{\theta}_{3,0} \tag{4.36}$$

$$\xi(t) = \hat{\tau}_0(t) - \hat{R}_s(t)\hat{\theta}_3(t-1) \tag{4.37}$$

$$k(t) = \frac{1}{\sigma_{e2}^2} Q(t)\hat{R}_s(t) \tag{4.38}$$

$$Q(t) = \frac{Q(t-1)\sigma_{e2}^2}{\sigma_{e2}^2 + \hat{R}_s^2(t)Q(t-1)} + \sigma_\omega^2, \quad Q(0) = \Phi_0 \tag{4.39}$$

where $\hat{\theta}_3(t)$ is the minimum-variance estimate of $C_0$ with Q(t) as its scalar error covariance.

### 4.3 Experimental Setup and Results

The subharmonic injection system overall diagram is shown in Fig. 4.4 where some parameters are already fixed due to the experimental setup but some others are required to be chosen. The fault resistors are connected to the stator windings taps via switches. The step-up and step-down transformer model the actual transmission and distribution lines.

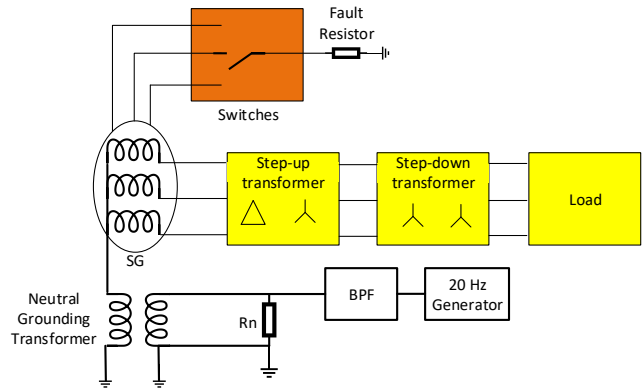

Fig. 4.4: 64S overall system block diagram.

A neutral grounding transformer connects the neutral circuit to the neutral node of the stator windings Y connection. This assures the high impedance grounding due to turns ratio and isolates the meaurment and injection circuit from the main synchronous machines. The power components of the set-up is shown below:

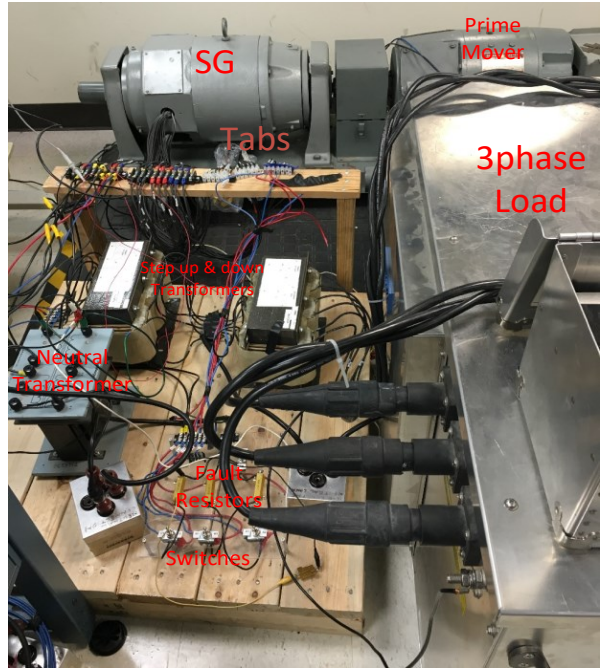

Fig. 4.5: Power circuit prototype.

The neutral resistor selection citeria was explained in the previous section. A 20Hz function generator supplies a subharmonic square wave which is passed through the band pass filter to inject a 20Hz sinusoidal voltage across the neutral resistor.

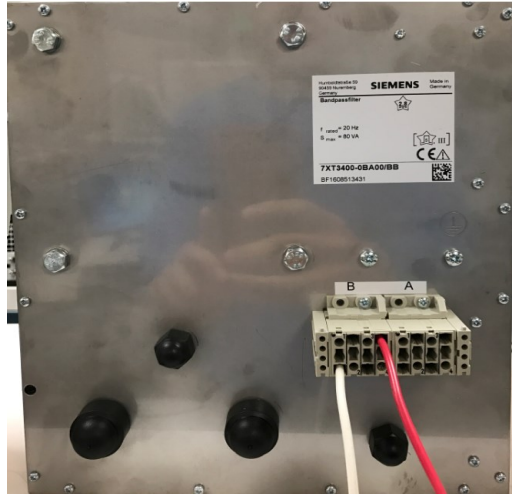

(a)

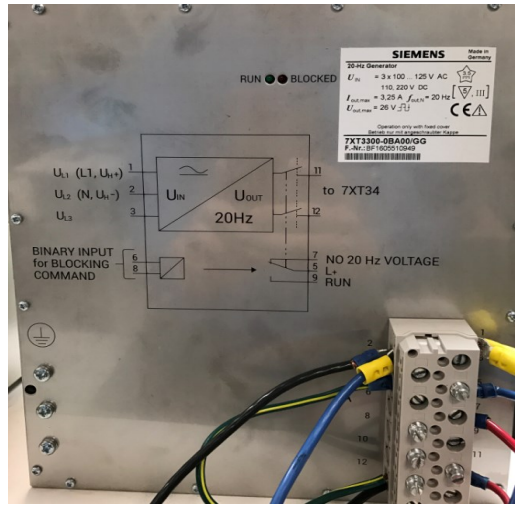

(b)

Fig. 4.6: Injection circuit components a) Band pass filter, b) 20Hz voltage generator.

These two elements of the injection circuit are shown in Fig. 4.6.

The sub-harmonic injection scheme responses for different fault scenarios are represented below. One of the distinctive features of the injection schemes compared to the third harmonic differential voltage scheme is the capability of detection when the generator is offline. Figs. 4.7 and 4.8 show the A64S estimations for hard terminal and hard neutral faults with 90Ω fault resistor. The insulation resistance ($R_s$) estimation approaches to 89Ω which is the actual insulation resistance after fault whereas 2.5kΩ insulation resistance is estimated for the healthy mode of operation. This reduction in the $R_s$ value can be used as a fault detection signature. The equivalent parasitic capacitance and time constant of the system is also depicted. Although detecting stator windings ground faults during offline mode is desirable, faults ocurring when the machine is online and is supplying power to the grid is of higher importance. Figs. (4.9-4.11) correspond to the online hard neutral faults at 90, 500 and 1000Ω fault resistors. The estimated $R_s$ also converges to the actual values. It should be stated that as fault resistor increases, the parallel combination of it and the real insulation resistance will diverge from $R_f$ such that $R_s$ is estimated to be 460Ω and 730Ω in Figs. 4.10 and 4.11, respectively. Figs. 4.12, 4.13 and 4.14 show the same responses when the fault is placed at 25% of the winding length. Figs. 4.15, 4.16 and 4.17 show the same responses when the fault is placed at 67% of the winding length. Finally Figs. 4.18, 4.19 and 4.20 belong to the hard terminal faults with the same fault resistors. Two rather rare cases are also analyzed in Figs. 4.21 and 4.22. During the generator stopping mode where the machine is speeding down and passes the 20Hz frequency, the main voltages and the injected sub-harmonic voltage may intersect. As shown in Fig. 4.21, a small perturbation in detection scheme happens and quickly it will recover itself to the correct values. As another less frequent case, when the machine opeartes at 600 rpm (20Hz) and a fault happens on the stator windings, the detction is not affected and the correct insulation resistances during both healthy and faulty opeartion is estimated.

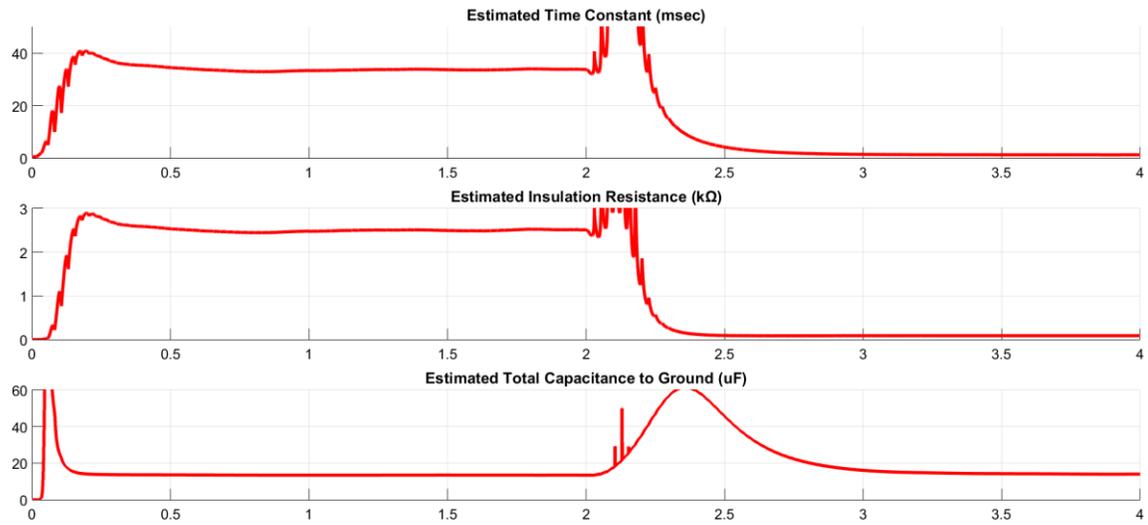

Fig. 4.7: A64S responses to the offline hard neutral fault with $R_f=90\Omega$.

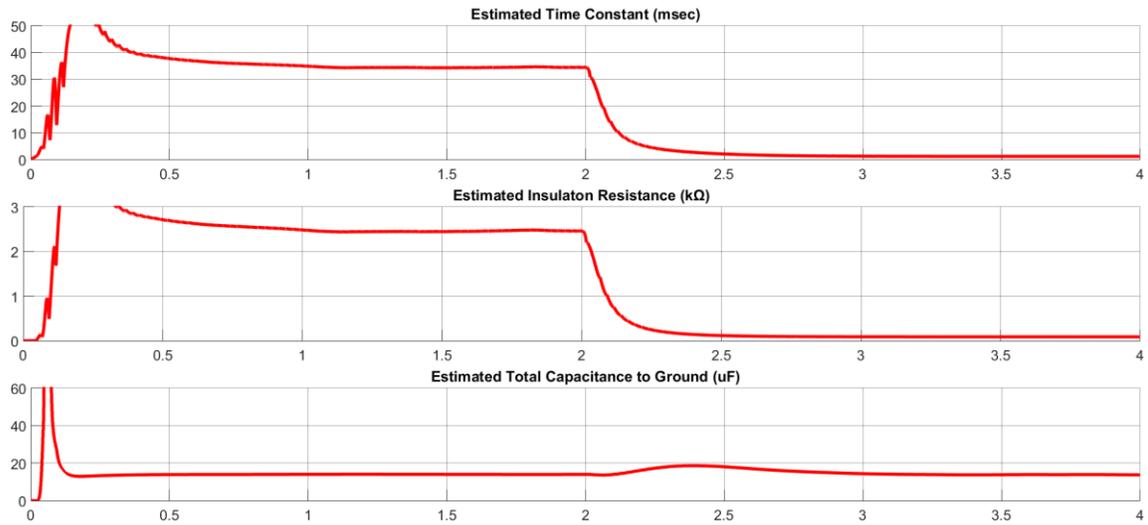

Fig. 4.8: A64S responses to the offline hard terminal fault with $R_f=90\Omega$.

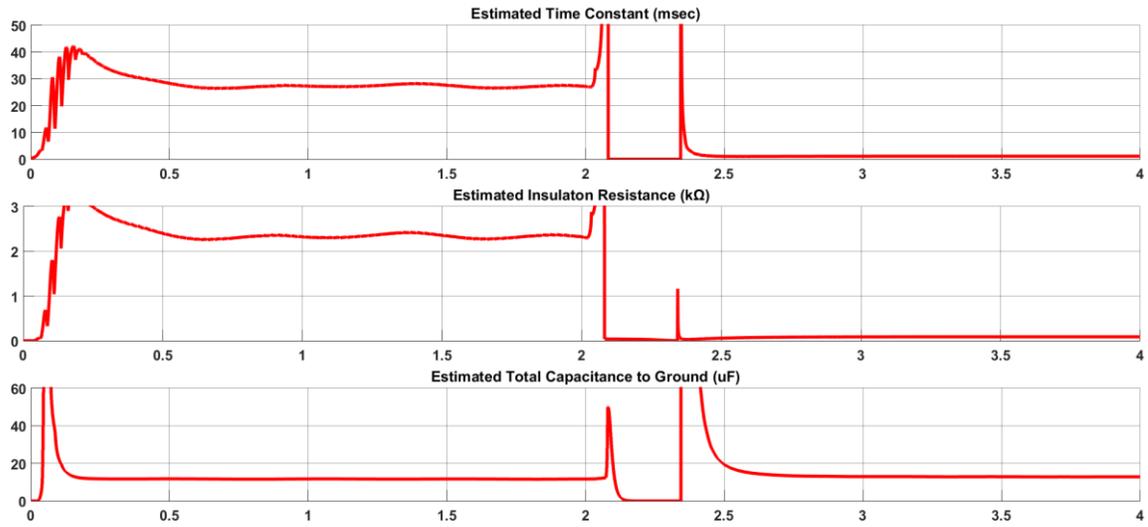

Fig. 4.9: A64S responses to the online hard neutral fault with $R_f$=90Ω.

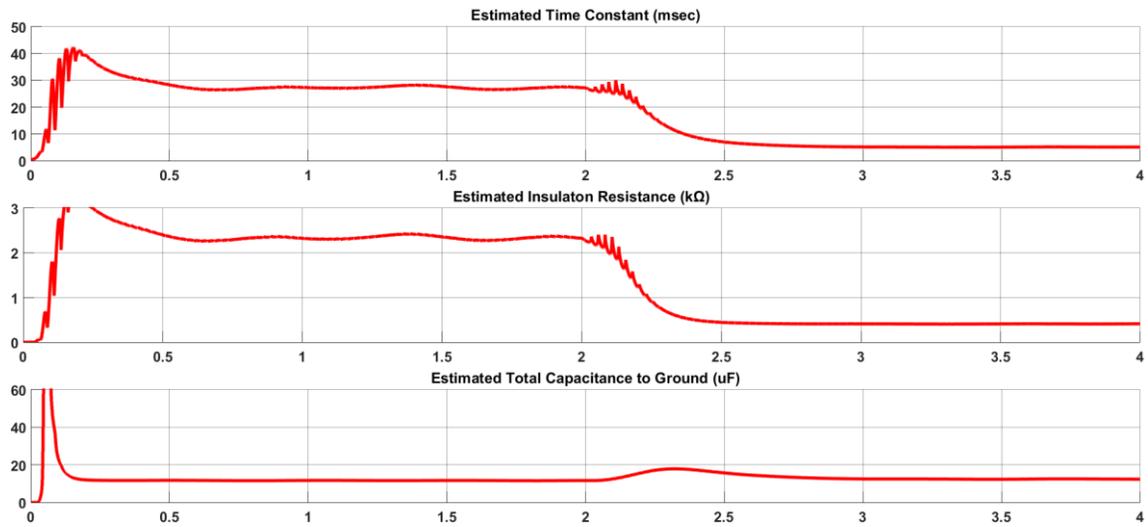

Fig. 4.10: A64S responses to the online hard neutral fault with $R_f$=500Ω.

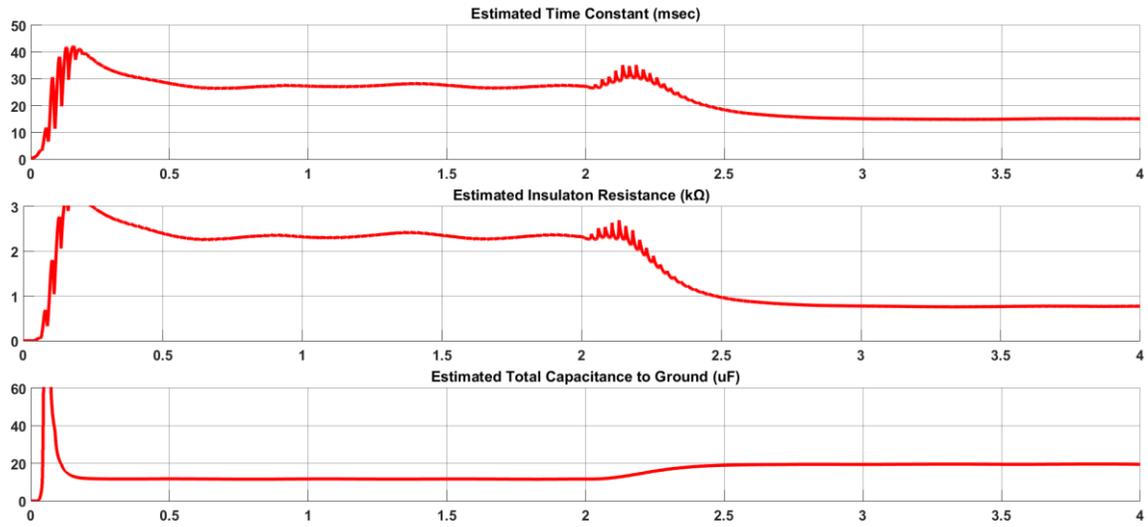

Fig. 4.11: A64S responses to the online hard neutral fault with $R_f$=1kΩ.

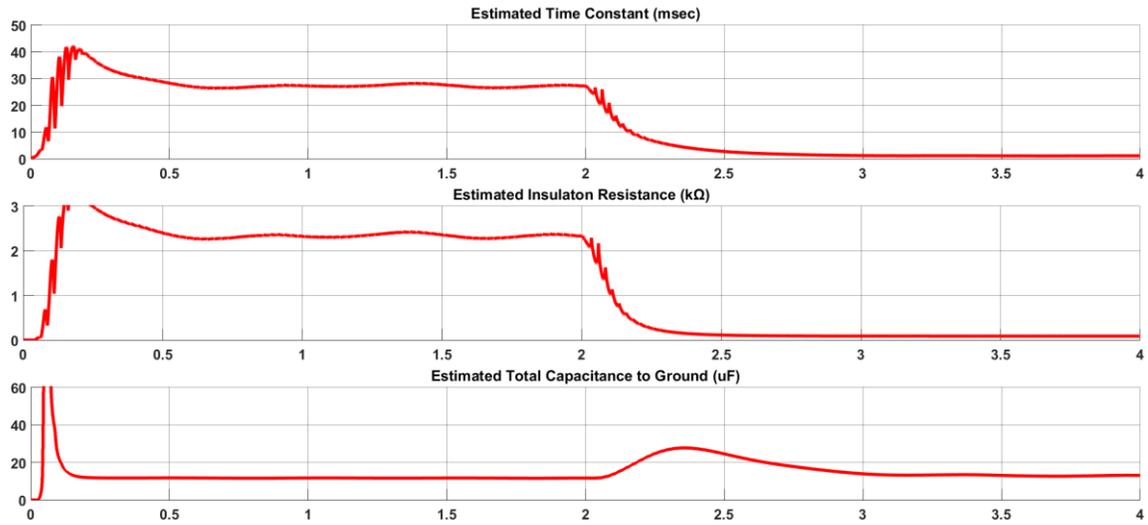

Fig. 4.12: A64S responses to the online 25% stator windings fault with $R_f$=90Ω.

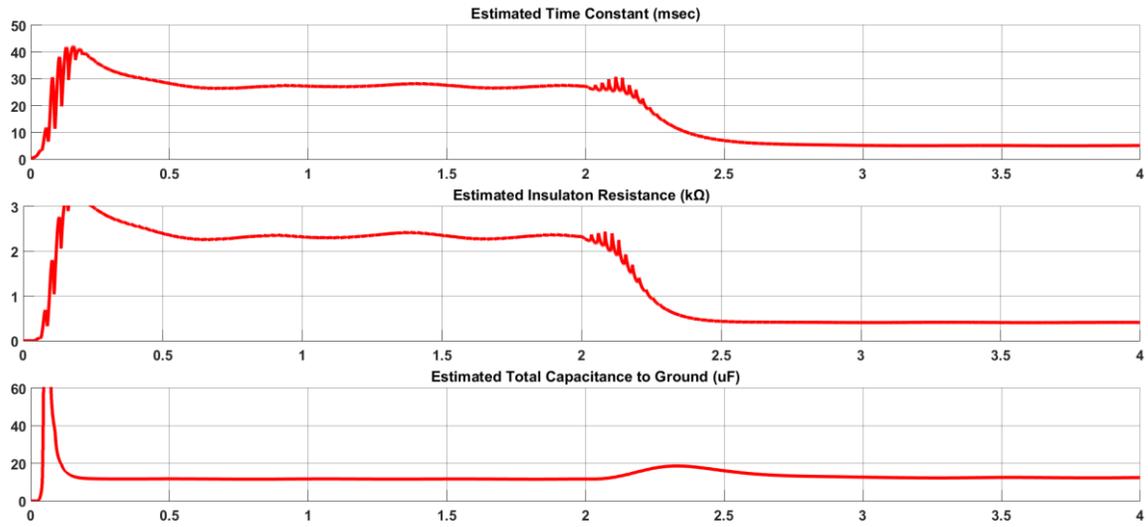

Fig. 4.13: A64S responses to the online 25% stator windings fault with $R_f=500\Omega$.

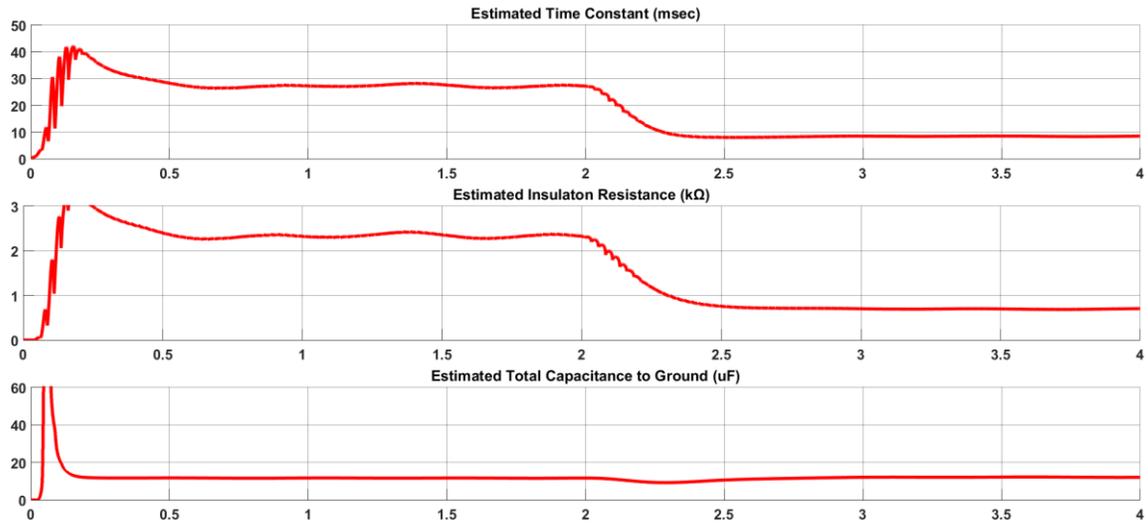

Fig. 4.14: A64S responses to the online 25% stator windings fault with $R_f=1k\Omega$.

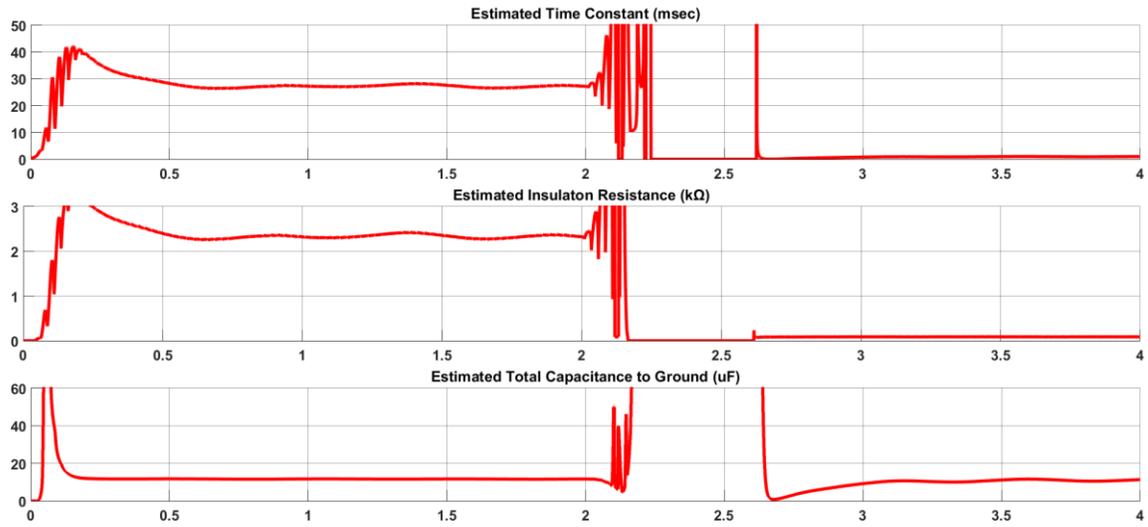

Fig. 4.15: A64S responses to the online 67% stator windings fault with $R_f=90\Omega$.

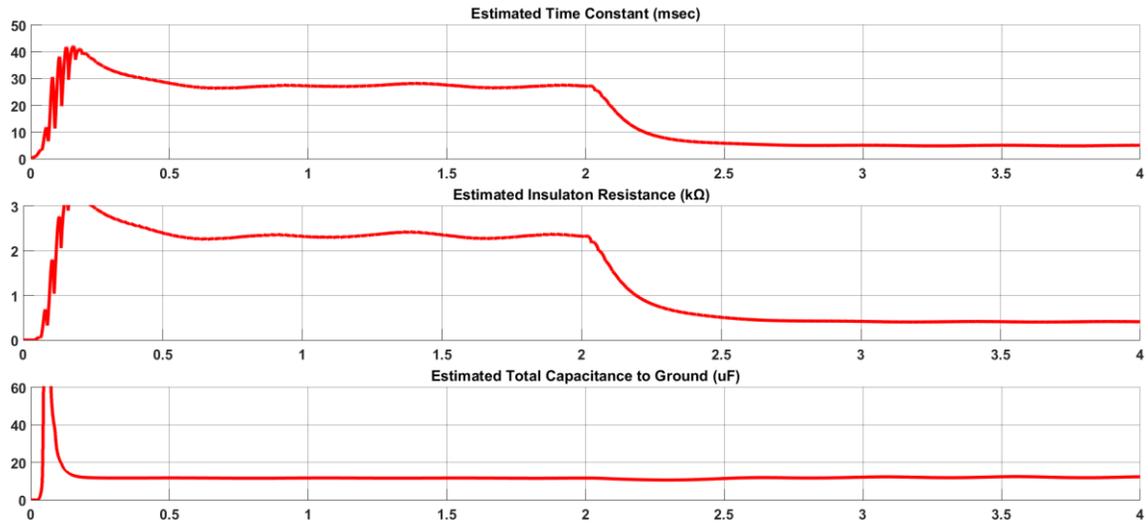

Fig. 4.16: A64S responses to the online 67% stator windings fault with $R_f=500\Omega$.

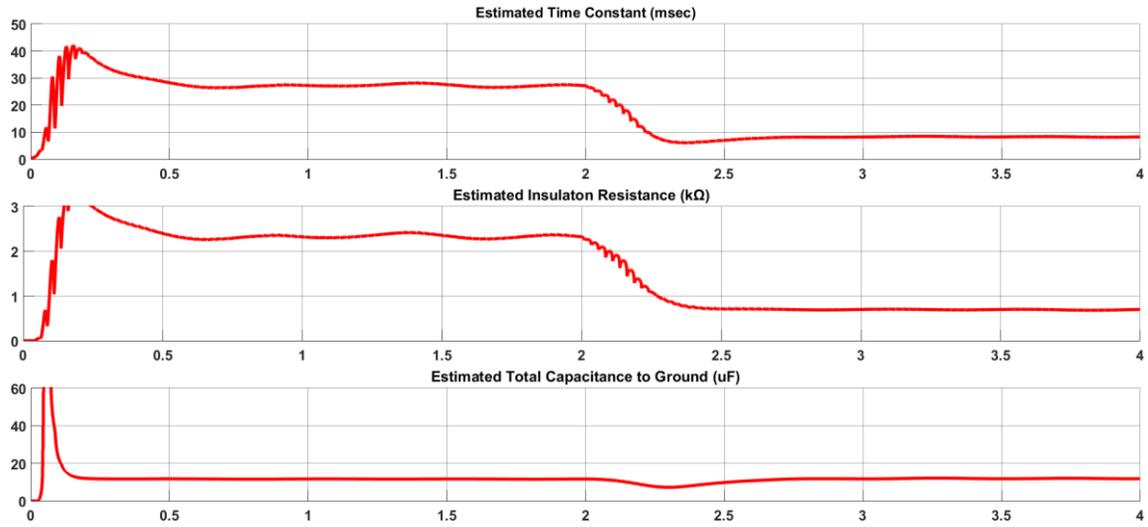

Fig. 4.17: A64S responses to the online 67% stator windings fault with $R_f$=1kΩ.

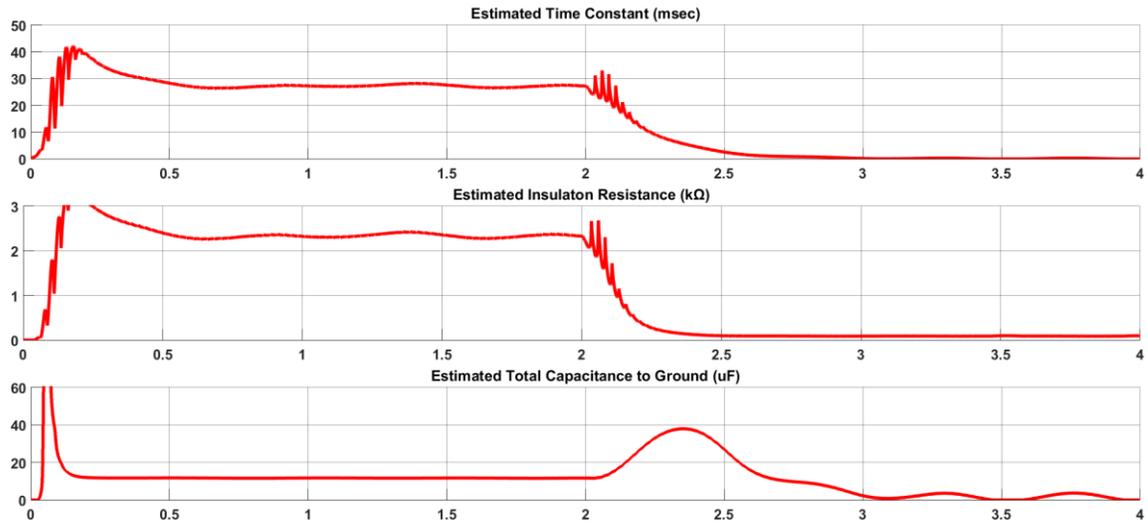

Fig. 4.18: A64S responses to the online hard terminal fault with $R_f$=90Ω.

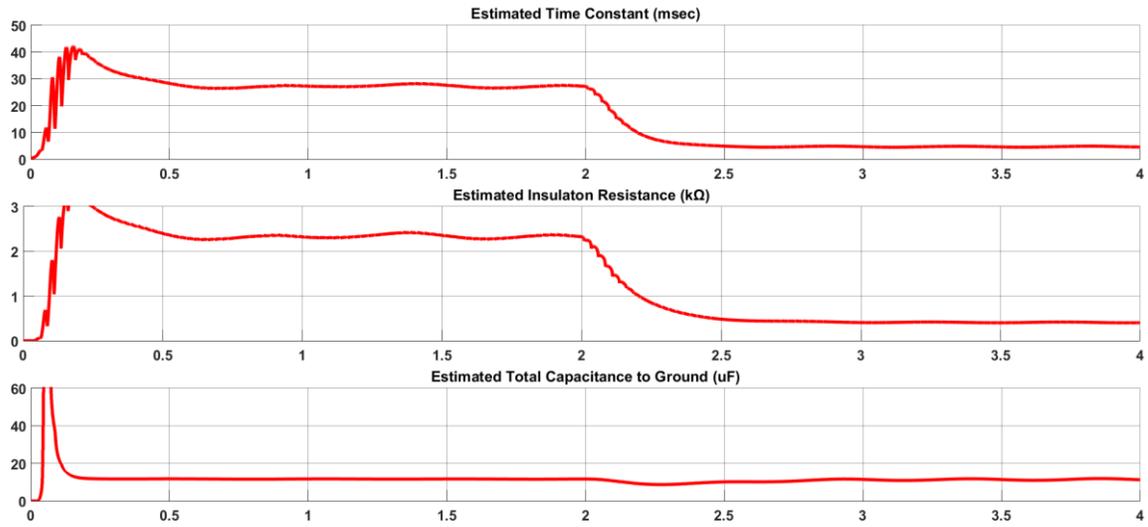

Fig. 4.19: A64S responses to the online hard terminal fault with $R_f=500\Omega$.

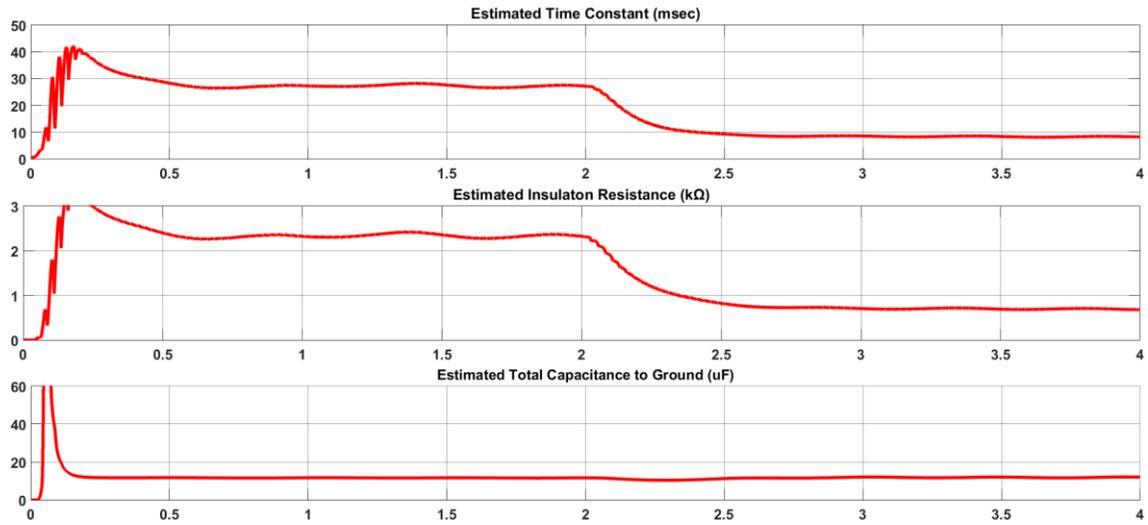

Fig. 4.20: A64S responses to the online hard terminal fault with $R_f=1k\Omega$.

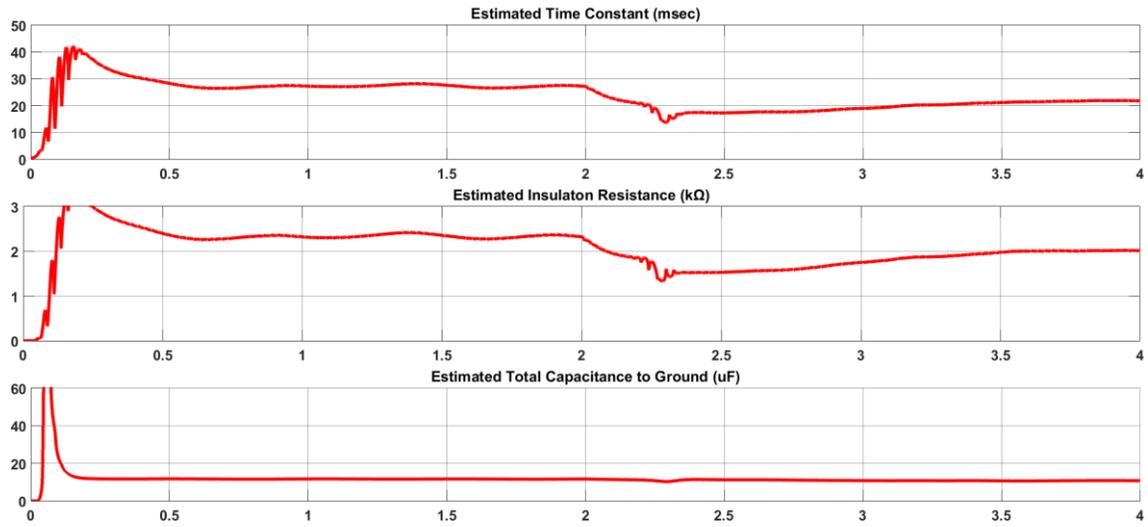

Fig. 4.21: A64S responses during generator stopping.

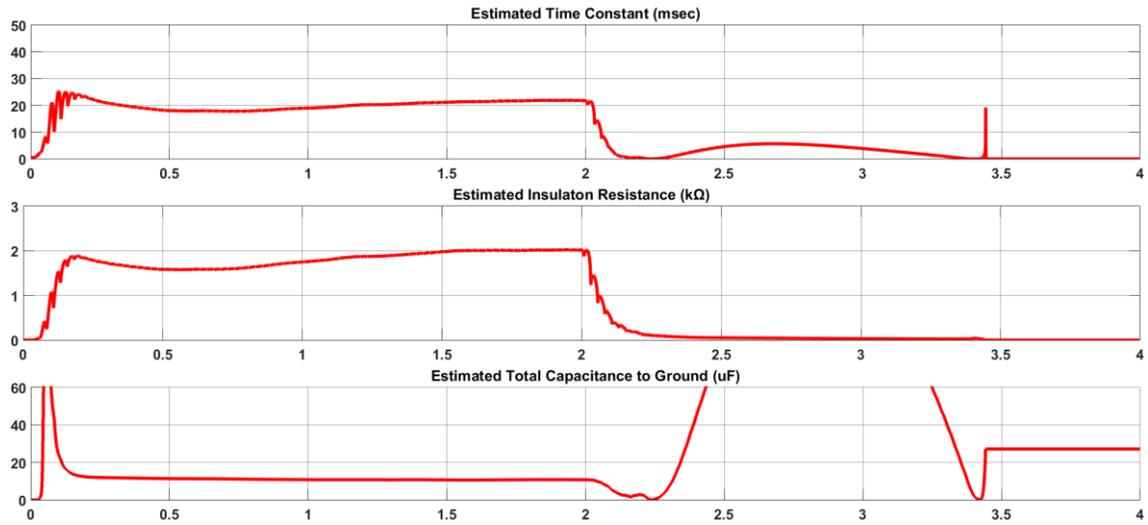

Fig. 4.22: A64S responses to the online hard terminal fault with $R_f$=90Ω at 600rpm.

## 4.4 Fault Location

As the 64S scheme was presented and the measurements and estimations are available through this scheme, a fault location detection scheme can be also developed to precisely determine the location of the faults along the winding. Reference [56] proposes a theoretical approach to calculate the fault location as shown below.

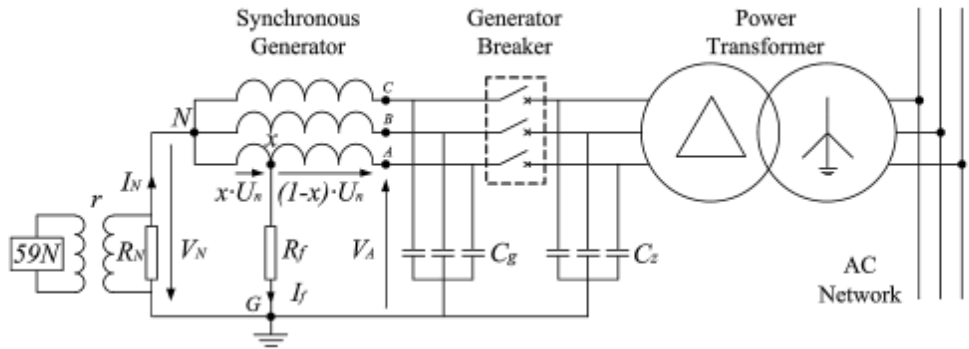

Fig. 4.23: Fault location scheme parameters & overal schematic.

The variable $x$ is the fault location where $x = 0$ is for a fault at the neutral shown as N and $x = 1$ is for a fault at the terminal. For fault at any other location, $0 < x < 1$. A formula is derived for computing $x$ based on the signals obtained from the 59N relay and the subharmonic injection scheme 64S:

$$x = \frac{V_N}{U_n R_N} \sqrt{(R_N + R_f)^2 + (\omega C_T R_f R_N)^2} \tag{4.40}$$

Here,

- $V_N$ is the 60 Hz component of the faulty neutral voltage in primary volts;
- $U_n$ is the rated 60 Hz line-to-ground voltage of the machine in primary volts;
- $C_T$ is the total capacitance to ground of the generator stator windings, iso-phase bus work and delta connected windings of the step-up transformer in uF;
- $R_N$ is the neutral grounding resistance in primary Ohms;

- $R_f$ is the stator single-line to ground fault resistance in primary Ohms;
- $\omega$ is the 60 Hz angular frequency, i.e., $\omega = 2\pi(60) = 377$ rad/sec.

Note that $x$ can be rewritten as:

$$x = \frac{V_N}{U_n R_N} \sqrt{(R_N + R_{Sf})^2 + (\omega R_N \tau_{0f})^2} \qquad (4.41)$$

where $\tau_{0f} = C_T R_{Sf}$ is the faulted stator insulation time-constant (msec) and $R_{Sf}$ is the faulted stator insulation resistance (kΩ).

The phase to neutral voltage (Un) is always constant regardless of the system operating either at healthy or faulty condition and the type of fault. Therefore, the fundamental frequency (60hz) component of the neutral voltage ($V_N$) in conjunction with the estimated insulation resistance, stator windings and connected apparatus equivalent capacitance can be used in fault location calculation.

Following figures demonstrate some results for different fault types. It should be mentioned that the arbitrary value of 2 is assigned to x during no fault operation

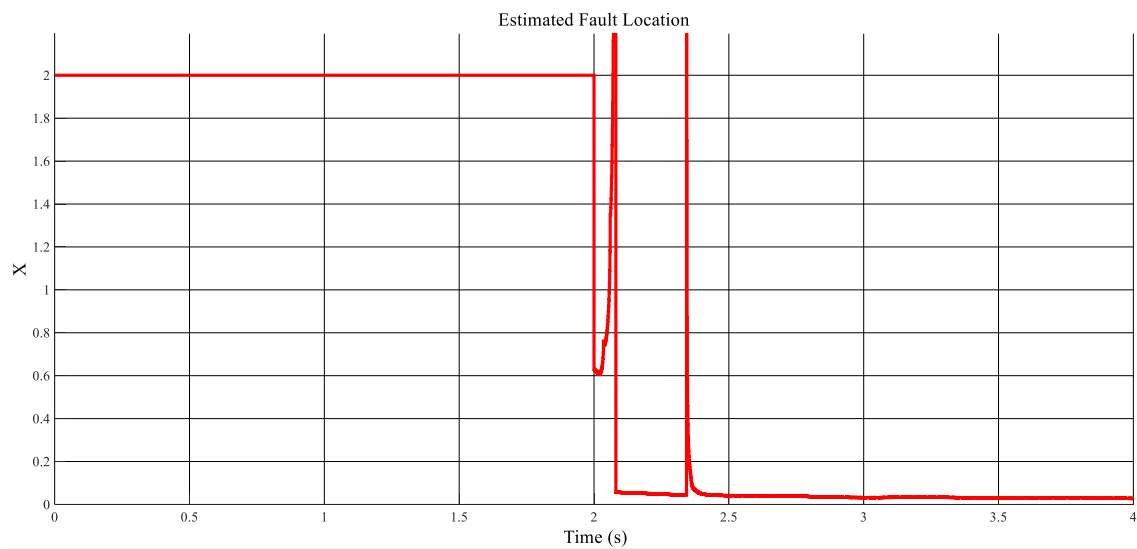

Fig. 4.24: Estimated fault location for hard neutral fault with $R_f$=90Ω (x=0.028).

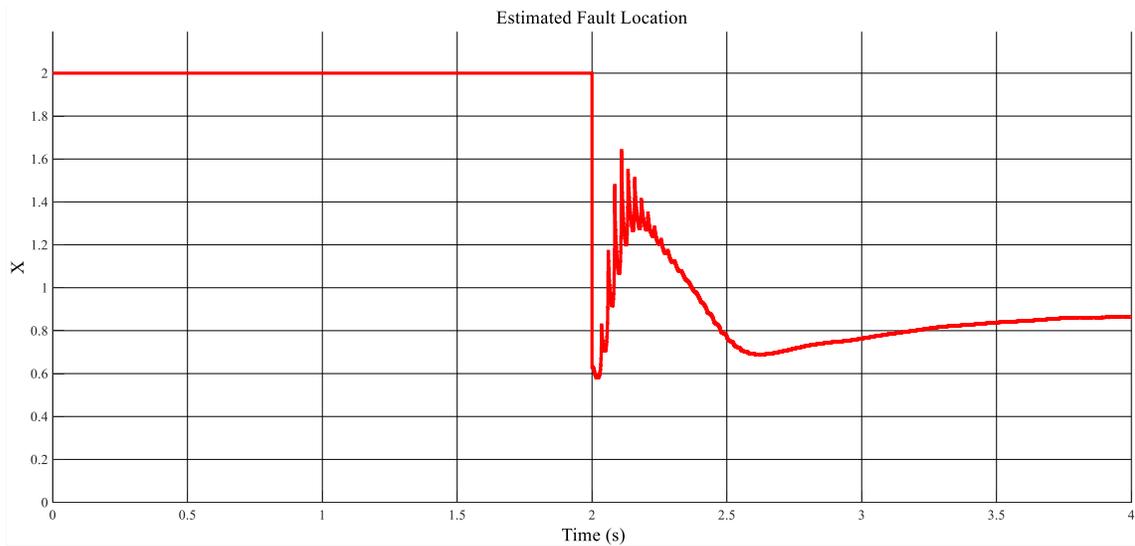

Fig. 4.25: Estimated fault location for hard terminal fault with $R_f$=90Ω (x=0.927).

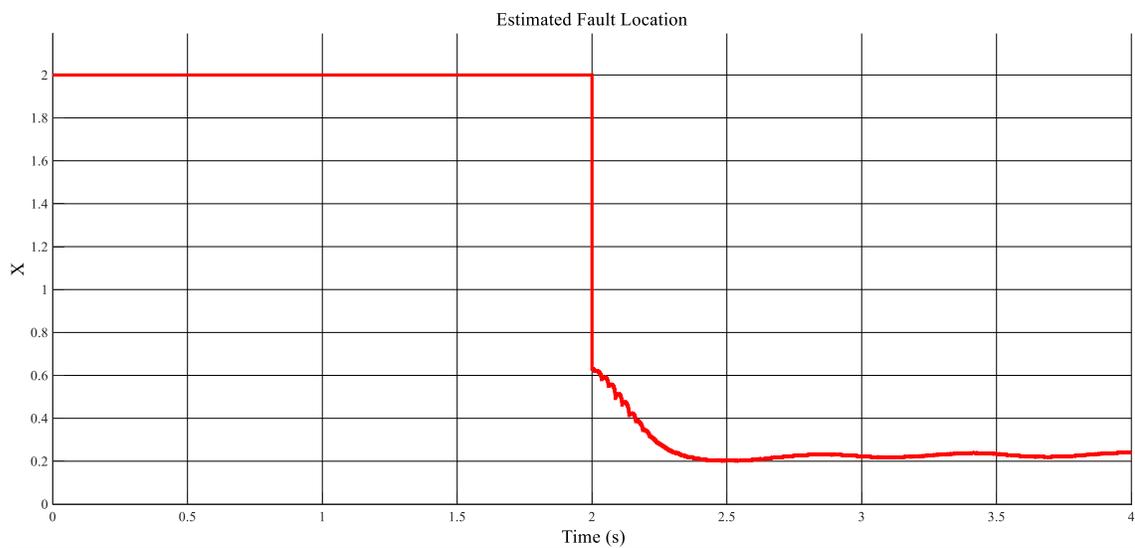

Fig. 4.26: Estimated fault location for 25% fault with $R_f$=1kΩ (x=0.25).

# 5 SUMMARY AND FUTURE WORK

The adaptive 3$^{rd}$ harmonic differential voltage scheme and the adaptive sub-harmonic injection scheme as two adaptive approaches for 100% stator windings ground fault protection in synchronous generators were discussed in this thesis.

The advantages of both adaptive schemes in comparison with non-adaptive schemes were also mentioned. Higher sensitivity and security in 3$^{rd}$ harmonic differential voltage scheme was shown and proved as A64G2 superiority versus 64G2 schemes. On the other hand, the adaptive sub-harmonic injection scheme offers higher sensitivity than the conventional injection schemes where overcurrent based schemes fail to detect high resistance faults and also some prior knowledge of the stator windings capacitor is required. The adaptive scheme can in fact estimate the total machine coupling capacitance and insulation resistance continuously and defines thresholds to make breaker trip decision.

The overall system including the machinery, power system components and data acquisition unit were modelled together to simulate the fault detection capability of the adaptive schemes. On the other hand, a comprehensive experimental setup was built to emulate the real generating stations as well as experimental implementation and verification of the adaptive schemes.

Thanks to the available experimental plant, several other fault scenarios can be modelled and analyzed. For example, arc modelling could be an interesting topic for future studies. As another instance, modifying the scheme for the cases where expected fault behavior for synchronous generators is not emulated could enhance the applicability of the adaptive scheme for the wider range of winding length.

The current adaptive scheme can be modified to apply to multi-generators connected to the same bus bar which is very prevalent in power plants.